\documentclass[hyper]{JHEP3}

\usepackage[english]{babel}
\usepackage{graphicx}
\usepackage{cite}
\usepackage{amsfonts}

\def\bea{\begin{eqnarray}}
\def\eea{\end{eqnarray}}
\newcommand{\la}{\left\langle}
\newcommand{\ra}{\right\rangle}
\newcommand{\aq}{\alpha_s\left( Q^2 \right)}
\newcommand{\lp}{\left(}
\newcommand{\rp}{\right)}
\newcommand{\GeV}{\mathrm{GeV}}
\def\l{\left}
\def\r{\right}

\title{The Bjorken sum rule with Monte Carlo and Neural Network
techniques}


\author{L.~Del~Debbio\\SUPA, School of Physics and Astronomy, University of Edinburgh\\            
  Edinburgh EH9 3JZ, Scotland\\ 
  Email:  \email{luigi.del.debbio@ed.ac.uk}}

\author{A.~Guffanti\\Physikalisches Institut, Albert-Ludwigs-Univerit\"at\\
  Hermann-Herder-Stra\ss e 3, 79104 Freiburg, Germany\\
  Email: \email{alberto.guffanti@physik.uni-freiburg.de}}

\author{A.~Piccione\\I.T.I.S. Pininfarina\\
  Via Ponchielli 16, 10024 Moncalieri, Italy\\
  and\\
  I.N.F.N. Milano\\
  Via Celoria 16, 20133 Milano, Italy\\
  Email: \email{hep@andreapiccione.it}}

\abstract{Determinations of structure functions and parton distribution 
  functions have been recently obtained using Monte Carlo methods and
  neural networks as universal, unbiased interpolants for the unknown
  functional dependence.  In this work the same methods are applied to
  obtain a parametrization of polarized Deep Inelastic Scattering
  (DIS) structure functions.  The Monte Carlo approach provides a
  bias--free determination of the probability measure in the space of
  structure functions, while retaining all the information on
  experimental errors and correlations. In particular the error on the
  data is propagated into an error on the structure functions that
  has a clear statistical meaning. We present the application of this
  method to the parametrization from polarized DIS data of the photon 
  asymmetries $A_1^p$ and $A_1^d$ from which we determine the structure 
  functions $g_1^p(x,Q^2)$ and $g_1^d(x,Q^2)$, and discuss the possibility 
  to extract physical parameters from these parametrizations. 
  This work can be used as a starting point for the determination of polarized 
  parton distributions.}

\preprint{FREIBURG-PHENO-09/04}

\begin{document}


\section{Introduction}
\label{sec:intro}

In QCD the description of scattering processes at large momentum transfer
($Q^2\gg\Lambda_{QCD}^2$) involving (polarized) hadrons in the initial state 
is based on the factorization theorem.  The latter allows a separation between the 
high--energy dynamics, described by coefficient
functions which are calculable in perturbative QCD, from low--energy,
non-perturbative effects, binding partons into hadrons, which are
encoded into (polarized) parton distribution functions (PDFs).

The growth in statistics and increase in precision of data from
experiments involving polarized hadrons scattering calls for a more
accurate determination of polarized PDFs and their errors.  A crucial
problem in this respect is the determination of the uncertainty on a
function (i.e.~a probability measure on a space of functions)
from a finite set of experimental data points. In the standard PDF extraction
approach to the problem the infinite--dimensional space of continuous functions 
is mapped into a finite--dimensional space of parameters by choosing a 
particular basis in the space of functions and truncating the basis to a finite 
number of elements. This procedure entails some degree of arbitrariness. 
Any sensible choice must strike a balance between two competing requirements: 
on the one hand a small number of parameters introduces a bias in the
determination of both the functional form and the errors, as the
chosen parametrization would not allow enough flexibility; on the
other hand a large number of parameters could spoil the convergence of
the fit, or be too sensitive to the statistical fluctuation of the
experimental data.

This problem has been addressed by the NNPDF Collaboration in the case
of unpolarized Deep Inelastic Scattering (DIS) structure functions in
Refs.~\cite{Forte:2002fg,DelDebbio:2004qj}, and in the case of the
unpolarized PDFs in Refs.~\cite{DelDebbio:2007ee,Ball:2008by,Ball:2009mk}
using a method based on statistical inference and neural networks as 
an interpolating tool.

While avoiding technical complications linked to the extraction of
PDFs from observables, the determination of structure functions
addresses the main issue of devising a faithful estimation of errors
on a function extracted from experimental data. The main ingredient in
the studies above is the usage of Monte Carlo methods to obtain a
representation of the probability measure in the space of structure
functions. An ensemble of artificial data is generated, which
reproduces all the statistical features (i.e.~variances and
correlations) of the original experimental data. Each set of
artificial data is called a replica. A structure function,
parametrized by a neural network, is then fitted to each replica. The
net result of this procedure is an ensemble of fitted functions. This
ensemble of fitted functions provides a representation of the measure
in the space of structure functions. Errors and correlations of any
observable involving the structure functions are obtained averaging
over the ensemble of fits. Moreover suitable statistical estimators
can be defined from the Monte Carlo ensemble which provide a
quantitative description of the possible biases and inconsistencies in
the fitting procedure. This method has been described in great detail
in Refs.~\cite{Forte:2002fg,DelDebbio:2004qj,DelDebbio:2007ee,Ball:2008by} to which the
interested reader should refer. 

The aim of this work is to apply the same techniques to obtain a
bias--free parametrization of the photon asymmetries $A_1^p$ and $A_1^d$ 
from available polarized DIS data and extract from them the corresponding
structure functions $g_1^p$ and $g_1^d$. We
provide further testing of the Monte Carlo method, and produce
statistically meaningful error bars for the structure
function. Besides allowing us to address all systematics related to
the data and the method, such a parametrization might be an ideal
input for a fit based on factorization scheme-invariant evolution
equations to determine $\alpha_s$, as proposed in
Refs.~\cite{Bluemlein:2002be,Bluemlein:2005rq}. As shown in this work,
a careful treatment of statistical and systematic errors leads to a
reliable extraction of physically meaningful parameters such as
$\alpha_s, g_A$, and the higher--twist contributions to the structure
functions. While these are not the best determinations available for these
parameters, the results we obtain are in agreement with other determinations,
and show the robustness of the Monte Carlo method.

We shall now discuss in turn the two steps that are needed to produce
the Monte Carlo sample of fitted functions: first the treatment of the
experimental data, and then the actual fitting procedure. The
experimental data points included in the fit are discussed in
Section~2; Section~3 summarizes briefly the NNPDF approach and the
characteristics of the neural networks used for this particular
study. The results of our fits, together with their phenomenological
implications are presented and discussed in Sections~4 and~5.

\section{Experimental data}
\label{sec:exp}

The cross section asymmetry for parallel and anti-parallel
configurations of longitudinal beam and target polarizations
is given by:
\bea
A_{||}=\frac{\sigma^{\uparrow\downarrow}-\sigma^{\uparrow\uparrow}}
{\sigma^{\uparrow\downarrow}+\sigma^{\uparrow\uparrow}}
\eea
and it is related to the virtual--photon asymmetries $A_1,A_2$ by:
\bea
A_{||}=D(A_1+\eta A_2)\simeq D A_1\, .
\label{eq:Apar}
\eea 

The photon depolarization factor $D$ depends on kinematic factors and
on the ratio:

\bea 
R(x, Q^2)=\frac{\sigma_L (x, Q^2)}{\sigma_T (x, Q^2)}\, , 
\eea
where $\sigma_L$ and $\sigma_T$ are the longitudinal and the
transverse cross sections respectively (see
e.g. Refs.~\cite{Anselmino:1994gn,Kuhn:2008sy} for a detailed
definition of all the quantities).
 
The polarized structure functions $g_1$ and $g_2$ are related to the
virtual-photon asymmetries by:
\bea
A_1 (x, Q^2)&=&\frac{g_1(x, Q^2)-\gamma^2 g_2(x, Q^2)}{F_1(x, Q^2)}\,,
\label{eq:A1}\\
A_2(x, Q^2)&=&\gamma\frac{g_1(x, Q^2)+g_2(x, Q^2)}{F_1(x, Q^2)}\,;
\label{eq:A2}
\eea
where
\bea
F_1(x,Q^2)=\frac{(1+\gamma^2)}{2x[1+R(x, Q^2)]}F_2 (x, Q^2)\,,
\label{eq:F1}
\eea
is the unpolarized structure function, $\gamma^2=\frac{4 m^2x^2}{Q^2}$, 
and $m$ denotes the nucleon mass.

The main features of each experimental data set used in the present analysis are summarized in
Tab.~\ref{tab:exp-data}, and their kinematical coverage of the $(x,Q^2)$-plane is shown in
Fig.~\ref{fig:kin_data}.  We observe that the kinematical coverage of the available data is rather
small, especially when compared to the one of the available unpolarized DIS data, thus we will have
a sizable region of the kinematical plane in which the fit extrapolates the behaviour extracted
from the region covered by data.

From Tab.~\ref{tab:exp-data} we infer that the systematic errors are on average 
one order of magnitude smaller than the statistical ones. This justifies the procedure
of neglecting correlations for systematic errors and the procedure of summing errors
in quadrature when computing the figure of merit ($\chi^2$) to be minimized in the fitting 
procedure.

Finally we notice that E155 data have been corrected to yield $A_1$ by adding in
Eq. (\ref{eq:A1}) the $g_2$ contribution evaluated with the Wandzura--Wilczek relation and 
using the parametrization of $g_1/F_1$ given in Ref.~\cite{Anthony:2000fn}:
this shift is also added as a source of uncertainty in the total error of the data set.

\TABLE
{\footnotesize
  \begin{tabular}{|c|c|c|c|c|c|c|c|c|c|} 
    \hline
    Experiment & $x$ range & $Q^2 (\GeV^2)$ range & $N_{\rm dat}$   
    & $\la\sigma_{\mathrm{stat}}\ra$ & $\la\sigma_{\mathrm{syst}}\ra$ 
    & $\la\sigma_{\mathrm{norm}}\ra$ 
    & Type & Ref.\\
    \hline
    \multicolumn{9}{|c|}{Proton}\\
    \hline
    EMC & 0.015 - 0.466 & 3.5 - 29.5 & 10 & 0.077 & 0.024 & 0.028 & 
    $A_1$ & \cite{Ashman:1989ig} \\
    \hline
    SMC & 0.001 - 0.480 & 0.3 - 58.0 & 15 & 0.026 & 0.003& 0.012 & 
    $A_1$ & \cite{Adeva:1998vv} \\
    \hline
    SMC low-$x$ & 0.0001 - 0.121 & 0.02 - 23.1 & 15 & 0.033 & 0.002 & 0.006 & 
    $A_1$ & \cite{Adeva:1999pa} \\
    \hline
    E143 & 0.031 - 0.75 & 1.27 - 9.52 & 28 & 0.045 & 0.016 & 0.012 & 
    $A_1$ &  \cite{Abe:1998wq}\\
    \hline
    E155 & 0.015 - 0.75 & 1.22 - 34.72 & 24 & 0.043 & 0.018 & 0.026 & 
    $g_1/F_1$ & \cite{Anthony:2000fn} \\
    \hline
    HERMES06 & 0.0058 - 0.7311 & 0.26 - 14.29 & 45 & 0.126 & 0.019 & 0.017 & 
    $A_1$ & \cite{Airapetian:2007mh} \\
    \hline
    \multicolumn{9}{|c|}{Deuteron}\\
    \hline
    COMPASS & 0.0051 - 0.474 & 1.18 - 47.5 & 12 & 0.034 & 0.017 & 0.011 & 
    $A_1$ & \cite{Ageev:2005gh} \\
    \hline
    SMC & 0.001 - 0.480 & 0.3 - 58.0 & 15 & 0.032 & 0.003 & 0.006 & 
    $A_1$ & \cite{Adeva:1998vv} \\
    \hline
    SMC low-$x$ & 0.0001 - 0.121 & 0.02 - 23.1 & 15 & 0.069 & 0.005 & 0.005 & 
    $A_1$ & \cite{Adeva:1999pa} \\
    \hline
    E143 & 0.031 - 0.75 & 1.27 - 9.52 & 28 & 0.066 & 0.011 & 0.008 & 
    $A_1$ & \cite{Abe:1998wq}\\
    \hline
    E155 & 0.015 - 0.75 & 1.22 - 34.72 & 24 & 0.091 & 0.009 & 0.011 & 
    $g_1/F_1$ & \cite{Anthony:2000fn} \\
    \hline
    HERMES06 & 0.0058 - 0.7311 & 0.26 - 14.29 & 45 & 0.089 & 0.007 & 0.009 & 
    $A_1$ & \cite{Airapetian:2007mh} \\
    \hline
  \end{tabular}
  {\caption{The proton and deuteron experimental data sets included in the present analysis. We
    show the kinematic range, the number of points, the average statistical, systematic and
    normalization uncertainty, and the measured observable.}}
  \label{tab:exp-data}
}

\FIGURE
{
  \includegraphics[scale=0.26]{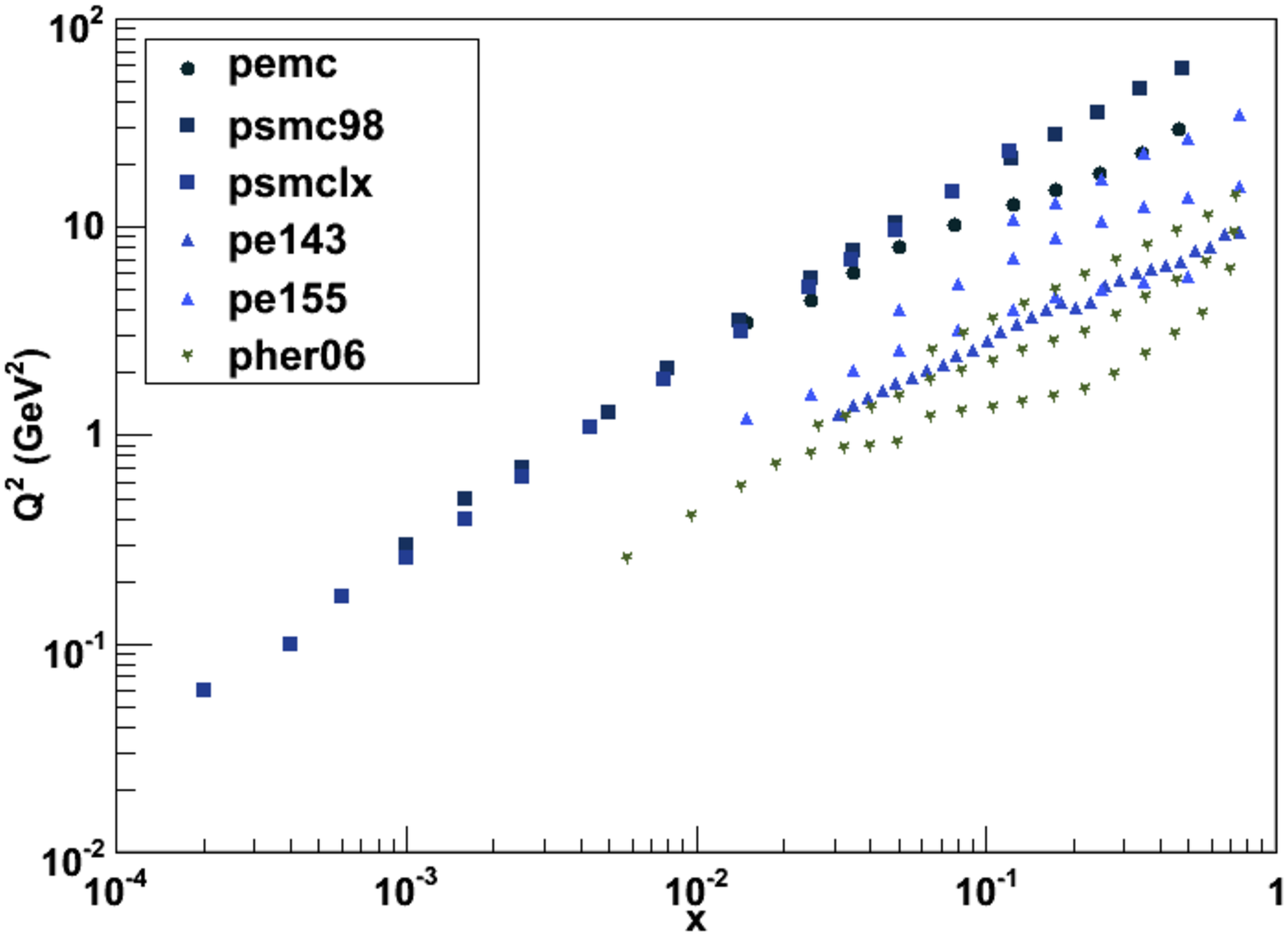}
  \includegraphics[scale=0.26]{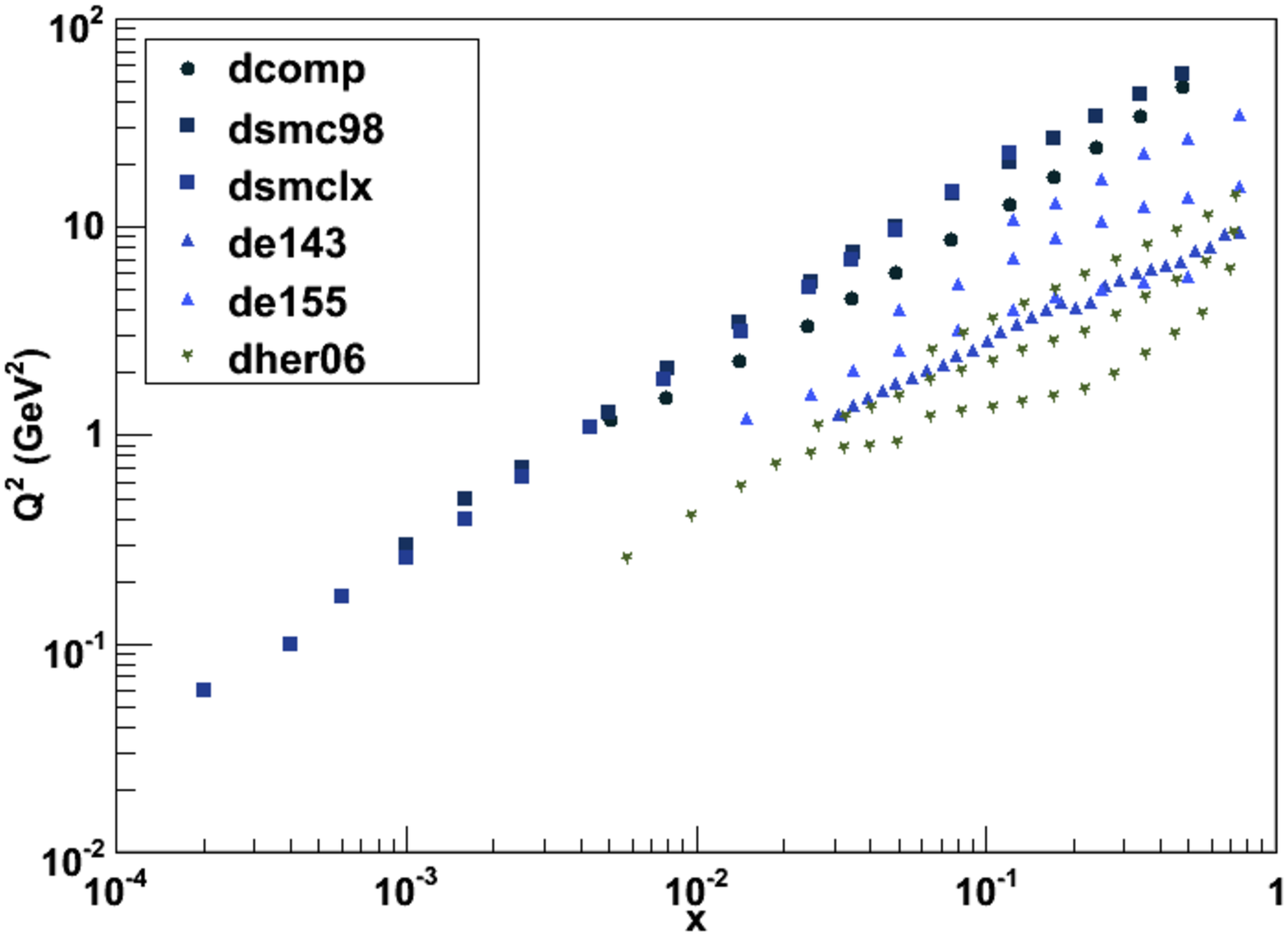}
  {\caption{Experimental data in the $(x,Q^2)$ plane used in the present analysis for the
    proton (left) and for the deuteron (right) target.}}
  \label{fig:kin_data}
}

\section{The NNPDF approach}
\label{sec:nnpdf}

In this section we briefly review the approach used to extract an
unbiased determination of the asymmetry $A_1$ and the structure
function $g_1$ from the available inclusive polarized DIS data,
following the analysis performed by the NNPDF Collaboration for the
determination of the unpolarized structure function $F_2$
\cite{Forte:2002fg,DelDebbio:2004qj} and the parton densities
\cite{DelDebbio:2007ee,Ball:2008by,Ball:2009mk}.

The core idea underlying the NNPDF approach is based on using Monte Carlo methods to build a
representation of the probability measure in the space of structure functions, and parametrizing
the space of structure functions using neural networks. We refer the interested reader to the
papers cited above for a detailed description of the methods and in the following we will briefly
discuss the settings used in this analysis. It is worthwhile emphasizing that the Monte Carlo
method does not require the use of neural networks, and would yield a robust determination of the
errors with any parametrization of the structure function, provided the parametrization is
sufficiently flexible. A comparison of Monte Carlo analyses based on different parametrizations was
performed in the framework of the HERA-LHC workshop by comparing the standard H1 and NNPDF
analyses. The greater flexibility of the neural network parametrization compared to fixed
functional forms is reflected in the larger error bands obtained using the NNPDF method, especially
when considering the $x$ region not covered by data (i.e. ~the extrapolation region). These
features are illustrated by the results in Sects.~3.2 and 3.4 of Ref.~\cite{Dittmar:2009ii}.

\subsection{Monte-Carlo replicas}

We generate $N_\mathrm{rep}$ Monte-Carlo replicas of the experimental
data according to
\begin{equation}
  A_1^{(\mathrm{art}),k} (x,Q^2)= \l(1+r_{k,N}
  \frac{\sigma_N}{A_1^{(\mathrm{exp})}(x,Q^2)}\r)
  \Big[A_1^{(\mathrm{exp})}(x,Q^2)+r_{k,t}\sigma_t(x,Q^2)\Big],
\end{equation}
where $r_k$ are Gaussian distributed random numbers, $\sigma_N$ is the
quadratic sum of the normalization errors and $\sigma_t$ is the total
error, obtained by summing in quadrature the statistical and
systematic errors, the latter assumed to be uncorrelated. 

Following Ref.~\cite{dagos}, the covariance matrix for experimental data points is evaluated using:
\begin{equation}
  {\rm cov}_{ij}=\sigma_{N_i}\sigma_{N_j}+\delta_{ij}\sigma_{i,t} \,,
\end{equation}
while during the fit for each replica, we minimize:
\begin{equation}
  \label{eq:chi2_min}
  \chi^{2 (k)}=\sum_{i=1}^{N_{\rm data}}\l(\frac{A_1^{(art),k}(x,Q^2) - A_1^{(net),k}(x,Q^2)}
  {\bar{\sigma}_{i,t}^{(k)}}\r)\,,
\end{equation}
where
\begin{equation}
  \bar{\sigma}_{i,t}^{(k)}=\l(1+r_{k,N}\frac{\sigma_N}{A_1^{(exp)}(x,Q^2)}\r)\sigma_{i,t}\,.
\end{equation}

The number of Monte Carlo replicas of the data is determined by
requiring that the average over the replicas reproduces the features
(central values, errors and correlations) of the original experimental
data to a required accuracy. The quantitative check is performed by
means of the statistical estimators described in the appendix of
Ref.~\cite{DelDebbio:2004qj} and the results for sets of 10, 100 and
1000 replicas are collected in Tab. \ref{tab:estimators_p} for the
proton target data and in Tab. \ref{tab:estimators_d} for the
deuteron target data. We observe that all the considered estimators 
have the correct scaling behaviour as the number of replica grows. 
We also point out that the large percentage error on the deuteron central 
values is due to a bulk of data whose values are close to zero.

\TABLE
{
  \begin{tabular}{|c|c|c|c|}
    \hline
    & 10 & 100 & 1000\\
    \hline
    $\langle PE\left[\langle A_1^{(art)}\rangle_\mathrm{rep}\right]\rangle$ & 14.20\%& 3.21\% &1.83\%\\ 
    $r\left[A_1^{(art)}\right]$ & 0.974203 & 0.998114 & 0.999682\\
    \hline
    $\langle V\left[\sigma^{(art)} \right]\rangle_{\mathrm{dat}}$& $3.1\cdot 10^{-3}$ & $1.5\cdot 10^{-3}$& $6.5\cdot 10^{-4}$\\
    $\langle PE\left[\langle \sigma^{(art)}\rangle\right]\rangle_{\mathrm{dat}}$ & 35.45\%& 12.44\% & 4.20\% \\ 
    $\langle\sigma^{(art)}\rangle_{\mathrm{dat}}$ & 0.0699 & 0.0766 & 0.0768  \\
    $r\left[\sigma^{(art)}\right]$ & 0.989956 & 0.997808 & 0.999793 \\
    \hline
    $\langle V\left[\rho^{(art)} \right]\rangle_{\mathrm{dat}}$& $9.5\cdot 10^{-2}$ & $8.9\cdot 10^{-3}$ & $9.2\cdot 10^{-4}$\\
    $\langle\rho^{(art)}\rangle_{\mathrm{dat}}$ & 0.1469 & 0.1567 & 0.1585  \\
    $r\left[\rho^{(art)}\right]$ & 0.638102 & 0.944682 & 0.993523  \\
    \hline
    $\langle V\left[ \mathrm{cov}^{(art)} \right]\rangle_{\mathrm{dat}}$& $7.6\cdot 10^{-5}$ & $1.1\cdot 10^{-5}$ & $1.5\cdot 10^{-6}$\\
    $\langle \mathrm{cov}^{(art)}\rangle_{\mathrm{dat}}$ & 0.00166 & 0.00154 & 0.00160  \\
    $r\left[ \mathrm{cov}^{(art)}\right]$ &  0.898803 & 0.986219  & 0.998858  \\
    \hline
  \end{tabular}
  {\caption{Statistical estimators for Monte Carlo replicas of $A_1$ for the 
    proton data. The experimental data have  
    $\langle\sigma^{(\mathrm{exp})}\rangle_{\mathrm{dat}}= 0.0764$,
    $\langle\rho^{(\mathrm{exp})}\rangle_{\mathrm{dat}}=0.1566$, and  
    $\langle \mathrm{cov}^{(\mathrm{exp})}\rangle_{\mathrm{dat}}= 0.00153$.}}
  \label{tab:estimators_p}
}

\TABLE
{
  \begin{tabular}{|c|c|c|c|}
    \hline
    & 10 & 100 & 1000\\
    \hline
    $\langle PE\left[\langle A_1^{(art)}\rangle_\mathrm{rep}\right]\rangle$& 89.71\%& 24.90\% & 5.97\%\\ 
    $r\left[A_1^{(art)}\right]$ & 0.977524 & 0.98633 & 0.999865 \\
    \hline
    $\langle V\left[\sigma^{(art)} \right]\rangle_{\mathrm{dat}}$& $5.3\cdot 10^{-3}$ & $1.7\cdot 10^{-3}$ & $6.7\cdot 10^{-4}$\\
    $\langle PE\left[\langle \sigma^{(art)}\rangle\right]\rangle_{\mathrm{dat}}$& 35.03\%& 11.75\% & 4.4\% \\ 
    $\langle\sigma^{(art)}\rangle_{\mathrm{dat}}$ & 0.0689 & 0.0739 & 0.0734 \\
    $r\left[\sigma^{(art)}\right]$ &  0.977501 & 0.997965 & 0.999705 \\
    \hline
    $\langle V\left[\rho^{(art)} \right]\rangle_{\mathrm{dat}}$& $1.0\cdot 10^{-2}$ & $9.2\cdot 10^{-3}$ & $8.7\cdot 10^{-4}$\\
    $\langle\rho^{(art)}\rangle_{\mathrm{dat}}$ & 0.0878 & 0.0904 & 0.0861  \\
    $r\left[\rho^{(art)}\right]$ & 0.612155 & 0.932158 & 0.992952 \\
    \hline
    $\langle V\left[ \mathrm{cov}^{(art)} \right]\rangle_{\mathrm{dat}}$& $6.3\cdot 10^{-5}$ & $1.0\cdot 10^{-5}$ & $1.0\cdot 10^{-6}$\\
    $\langle \mathrm{cov}^{(art)}\rangle_{\mathrm{dat}}$ & 0.00133 & 0.00145 & 0.00134 \\
    $r\left[ \mathrm{cov}^{(art)}\right]$ &  0.959275 & 0.995339 & 0.999479 \\
    \hline
  \end{tabular}
  {\caption{Statistical estimators for Monte Carlo replicas of $A_1$ for 
    the deuteron data. The experimental data have  
    $\langle\sigma^{(\mathrm{exp})}\rangle_{\mathrm{dat}}= 0.0733$,
    $\langle\rho^{(\mathrm{exp})}\rangle_{\mathrm{dat}}=0.0862$, and  
    $\langle \mathrm{cov}^{(\mathrm{exp})}\rangle_{\mathrm{dat}}= 0.00135$.}}
  \label{tab:estimators_d}
}

\subsection{Neural Networks as unbiased interpolants}

Artificial neural networks, see e.g. Ref.~\cite{NN}, are a class
of algorithms which provide a robust and universal approximant to
incomplete or noisy data, with the only requirement of continuity. 
Neural networks are universal approximators for measurable functions~\cite{approx}.
This means that any continuous function can be approximated to 
any degree of accuracy by a sufficiently large neural network with one 
hidden layer and non-linear neuron activation function.

One of the main reasons to use neural networks in place of any other redundant parametrization is
the existence of efficient techniques for {\em training} them, i.e.~determining the parameters of
the network (thresholds and weights) so that it reproduces a given set of input-output data.
Equivalently one could say that a sufficiently large neural network provides a description of the
data which is largely free of functional bias.

The analysis presented here uses a class of neural networks known as multilayer feed-forward
perceptrons, trained using a genetic algorithm \cite{genetic,LR}.  The networks we employed have
one hidden layer and a 2-4-1 architecture, which gives us a total of 17 free parameters for each
network to be determined during the training. The guidance principle in the choice of the network
architecture to be used is that it should provide a redundant parametrization for the data to be
fitted, i.e.~the network should have enough flexibility to fit not only the underlying physical
law but also the statistical fluctuations of the experimental data.  This property is crucial in
ensuring that the fit results are not biased by the specific parametrization. The lack of
functional bias is established a posteriori by verifying that fits performed with networks with
different architectures lead to statistically equivalent results. This is achieved using the
statistical estimators introduced in the NNPDF Collaboration's studies; the results of these
comparisons are presented and discussed later.

\FIGURE
{
  \includegraphics[scale=0.3]{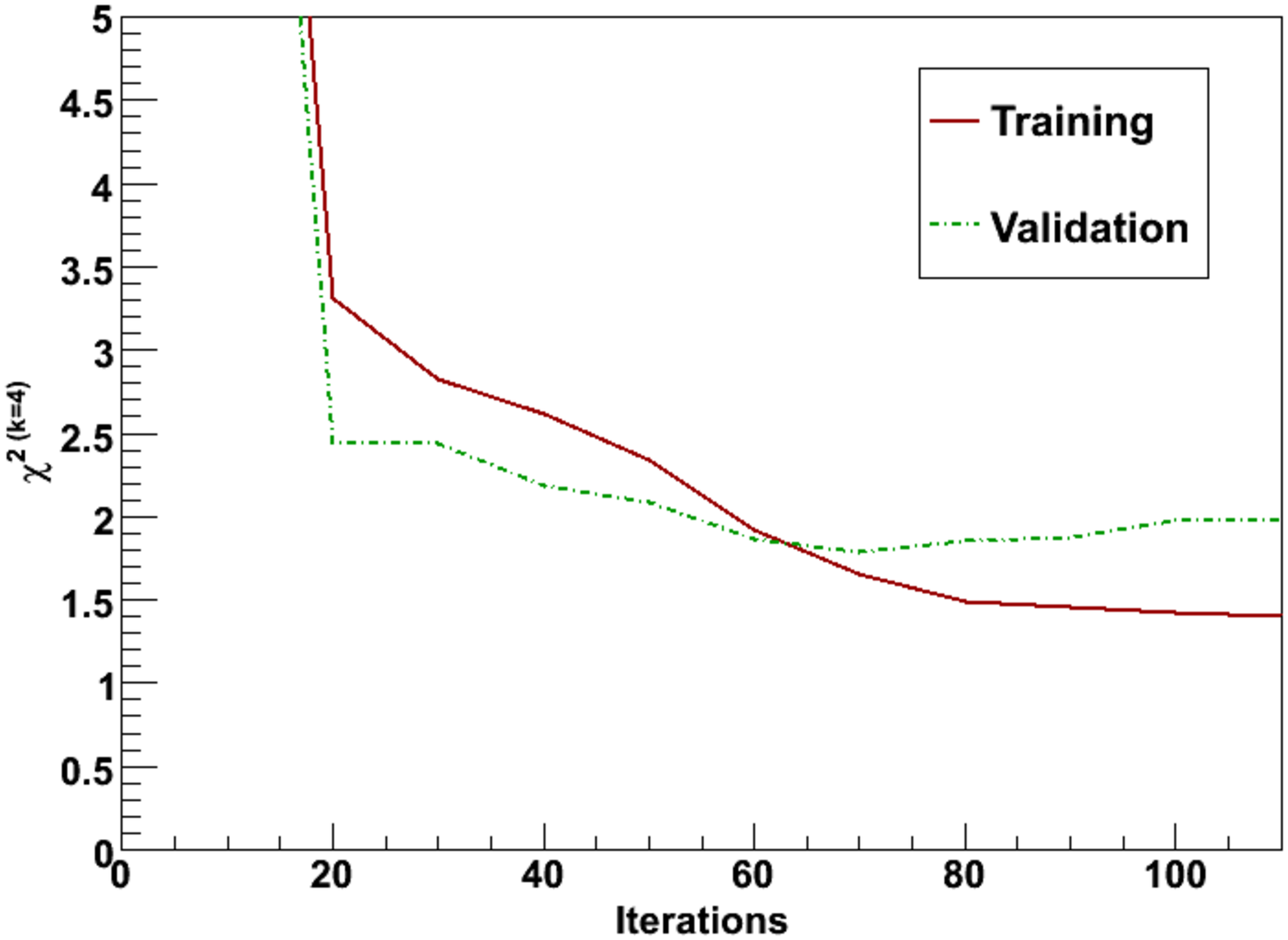}
  {\caption {$\chi^2$ for the training (red) and validation (green) sets of one replica 
    in the reference fit to $A_1^p$.}}
  \label{fig:tr-val-chi2}
} 
 
The training of the individual networks to the Monte Carlo replicas is
performed by minimizing the figure of merit given in
Eq.~(\ref{eq:chi2_min}). Given the extensive size and complex
structure of the parameter space (a neural network with $n$
parameters, weights and thresholds has $2n!$ equivalent global minimum
configurations), the most efficient training algorithm turns out to be
a genetic algorithm. The details of the implementation are discussed
in Ref.~\cite{DelDebbio:2007ee}.

As already pointed out in various references the fact that we adopt a
redundant parametrization and that the figure of merit minimized in 
the training procedure is monotonically decreasing might lead to 
\emph{overfitting} the data: the neural network reproduces not only the underlying 
physical law but also the statistical noise of the data sample. To prevent this 
from happening and to determine the optimal fit we adopt a criterion to stop our fit based 
on the \emph{cross-validation} method. Once again our procedure is completely
analogous to the one used for the unpolarized NNPDF fits.

For each replica of the experimental data we subdivide the data into a
\emph{training} and a \emph{validation} set, respectively containing a
fraction $f_\mathrm{tr}$ and $(1-f_\mathrm{tr})$ of randomly chosen
data points of each experiment.

We train one neural network on each replica of the data using the $\chi^2$ of 
the training set as a figure of merit to be minimized. In parallel we compute
the $\chi^2$ of the validation set.
We stop the training when we find that the $\chi^2$ smeared over a given number
of generations is decreasing for the training set while increasing for the 
validation set.
A graphical illustration of such a behaviour for one of the replicas in the 
reference fit is given in Fig. \ref{fig:tr-val-chi2}.

\section{Phenomenology}
\label{sec:pheno}

The study of the first moments of polarized structure functions is of
phenomenological interest, since they can be used to extract
information on the fraction of polarization carried by partons and on
physical couplings. In the $\overline{\mathrm{MS}}$ scheme we have 

\begin{eqnarray}
  \Gamma_1^{p,n} (Q^2)&=&\int_0^1 dx\,g_1^{p,n} (x, Q^2)= \\ \nonumber
  &=&\frac{1}{36}\left[(a_8\pm3a_3)\Delta C_{\mathrm{NS}}^{\overline{\mathrm{MS}}}
  (\alpha_s(Q^2)) +4a_0\Delta C_{\mathrm{S}}^{\overline{\mathrm{MS}}} (\alpha_s(Q^2))
  \right]\, ,
\end{eqnarray}
where $\Delta C_{\mathrm{NS}}^{\overline{\mathrm{MS}}} (\alpha_s(Q^2))$ and $\Delta
C_{\mathrm{S}}^{\overline{\mathrm{MS}}} (\alpha_s(Q^2))$ are the first moments of the
non-singlet and singlet Wilson coefficient functions, respectively,
and 
\begin{eqnarray}
  a_3&=&(\Delta u + \Delta \bar{u})-(\Delta d + \Delta \bar{d})\,,\\
  a_8&=&(\Delta u + \Delta \bar{u})+(\Delta d + \Delta \bar{d})-
  2(\Delta s + \Delta \bar{s})\,,\\
  a_0&=&(\Delta u + \Delta \bar{u})+(\Delta d + \Delta \bar{d})+ (\Delta
  s + \Delta \bar{s})\equiv\Delta\Sigma\,.  
\end{eqnarray}

Using isotopic spin invariance, it can be shown that $a_3$ is the
axial coupling $g_A=G_A/G_V$ that governs neutron
$\beta$-decay. Accurate measurements yield (see e.g. Ref.~\cite{Kuhn:2008sy}):
\begin{equation}
  g_A=1.2670\pm0.0035\, .  
\end{equation}
 
The difference of the $g_1$ moments for proton and neutron leads to
the Bjorken sum rule  
\begin{equation}
  \Gamma_1^{\mathrm{NS}}(Q^2)=\Gamma_1^p(Q^2)-\Gamma_1^n(Q^2)=\frac{1}{6}g_A
  \Delta C_{\mathrm{NS}}^{\overline{\mathrm{MS}}} (\alpha_s(Q^2))+\delta_T+\delta_\tau\,;
\label{eq:bjsr}
\end{equation}
where $\delta_T$ is the target mass correction and $\delta_t$ is the
correction due to higher twists.  Target mass corrections have been
studied in Refs.~\cite{Matsuda:1979ad,Piccione:1997zh,Blumlein:1998nv,Accardi:2008pc} 
and can be evaluated for any moment $n$ at the first order in $m^2/Q^2$ using
\cite{Piccione:1997zh}:
\begin{equation}
  \delta_T=g_1^{(n)}
  (Q^2)-g_{10}^{(n)}(Q^2)=\frac{m^2}{Q^2}\frac{n(n+1)}{(n+2)^2}
  \l[(n+4)g_{10}^{n+2}(Q^2)+4\frac{n+2}{n+1}g_{20}^{n+2}(Q^2)\r]\,,
  \label{eq:tmc}
\end{equation}
where $g_i^{(n)}(Q^2)=\int_0^1 dx\,x^{n-1} g_i(x,Q^2)$ and $g_{i0}$ is
the structure function taken at zero mass of the nucleon.  The
higher--twist contribution is simply
\begin{equation}
  \delta_\tau=\frac{\mu_4}{Q^2}\,,
  \label{eq:f2_HT}
\end{equation}
where $\mu_4$ can be extracted from experimental data at low $Q^2$
such as the CLAS data~\cite{Simolo:2006iw}.  Finally, the coefficient
function of Eq.~(\ref{eq:bjsr}) has been calculated in
Ref.~\cite{Larin:1991tj} and up to order $\alpha_s^3$ is given by:
\begin{eqnarray}
  \Delta C_{\mathrm{NS}}^{\overline{\mathrm{MS}}} (\alpha_s(Q^2))&=&1-\frac{\alpha_s
    (Q^2)}{\pi}- \left(\frac{55}{12}-\frac{n_f}{3}\right)
  \left(\frac{\alpha_s (Q^2)}{\pi}\right)^2
  \nonumber\\
  &-& \left(41.4399-7.6072\,n_f +\frac{115}{648}\,n_f^2\right)
  \left(\frac{\alpha_s (Q^2)}{\pi}\right)^3\, .  
\end{eqnarray}
For the running coupling we use the expanded solution of the
renormalization group equation, up to NNLO we have:
\begin{eqnarray}
  \aq&=&\aq_{\mathrm{LO}}\Bigg[ 1+\aq_{\mathrm{LO}}\left[ \aq_{\mathrm{LO}}- \alpha_s\lp M_Z^2\rp
  \right]
  (b_2-b_1^2) \nonumber \\
  &+&\aq_{\mathrm{NLO}}b_1 \ln\frac{\aq_{\mathrm{NLO}}}{\alpha_s\lp M_Z^2\rp}\Bigg], 
\end{eqnarray}
with 
\begin{eqnarray} 
  \aq_{\mathrm{NLO}}=
  \aq_{\mathrm{LO}}
  \left[ 1-b_1\aq_{\mathrm{LO}}\ln\lp 1+\beta_0\alpha_s\lp M_Z^2\rp \ln\frac{Q^2}{M_Z^2}\rp\right] \ ,
\end{eqnarray} 
\begin{eqnarray} 
  \aq_{\mathrm{LO}}=\frac{\alpha_s\lp M_Z^2\rp}{ 1+\beta_0\alpha_s\lp
    M_Z^2\rp \ln\frac{Q^2}{M_Z^2}}, 
\end{eqnarray} 
and the beta function coefficients given by 
\begin{eqnarray} 
  Q^2\frac{da_s(Q^2)}{d Q^2}=-\sum_{k=0}^2
  \beta_k a_s(Q^2)^{k+2} , \qquad a_s(Q^2)=\frac{\alpha_s(Q^2)}{4\pi}
  \, , 
\end{eqnarray}
where 
\begin{eqnarray}
  \beta_0 &=& 11-\frac{2}{3}n_f \ , \\
  \beta_1 &=& 102-\frac{38}{3}n_f \ ,\nonumber\\
  \beta_2 &=& \frac{2857}{2}-\frac{5033}{18}n_f+\frac{325}{54}n_f^2 \,,\nonumber
\end{eqnarray}
and $b_i\equiv \beta_i/\beta_0$.

\section{Results}
\label{sec:res}

In this section we present our parametrization of the proton and deuteron
asymmetries and the structure functions extracted from them.
 
We assess the quality of the fit by comparing our extraction with the
experimental data included in the analysis, and by studying the stability of our 
results under variations of the parametrization used for the networks. 

Then, as an example of a possible application of our result to a phenomenological analysis, we
study the extraction of the physical parameters (the strong coupling constant $\alpha_s$ and the
axial coupling $g_A$) from the Bjorken sum rule.  In order to give a faithful error on the
extracted quantities we study the impact of the different assumptions which are needed to
reconstruct the structure functions and then to evaluate the Bjorken sum rule. Results are compared
to existing estimates.

\subsection{The final fit and its statistical features}

\TABLE
{
    \begin{tabular}{|c|c||c|c|} 
      \hline
      Proton & $\chi^2$ &  Deuteron & $\chi^2$ \\
      \hline
      \hline
      EMC & 0.370 & COMPASS & 0.885\\
      SMC & 0.480 & SMC &  1.100 \\
      SMC low-$x$ & 1.150 & SMC low-$x$ & 0.774 \\
      E143 & 0.904 & E143 & 1.530 \\
      E155 & 0.717 & E155 & 0.661 \\
      HERMES06 & 0.456 & HERMES06 & 0.881\\
      \hline
      Total & 0.666 & Total & 0.986 \\
      \hline
    \end{tabular}
  {\caption{The $\chi^2$ of the fit for proton and deuteron data.}}
  \label{tab:pd_chi2}
}

In Tab.~\ref{tab:pd_chi2} we show the $\chi^2/N_\mathrm{data}$ for
each target and each experimental data set included in the present analysis. 
We first observe the overall good quality of our fit. 
For the proton the small values of $\chi^2$ for EMC, SMC and HERMES can 
be explained by a possible overestimate of experimental errors. 
For the deuteron all the $\chi^2$ are of order 1, except for the E155 
data set which has a value of $\chi^2$ significantly smaller than one.
The somewhat larger value of $\chi^2$ for the E143 deuteron data set 
can be understood by looking at Figs.~\ref{fig:fit_vs_data_1} and~\ref{fig:fit_vs_data_2} 
where we present a comparison of our fit to experimental data in different 
kinematical regions.  We observe that in the case of E143 the deuteron 
data show small incompatibilities among themselves, and the large value of $\chi^2$ 
is a reflection of this. It is interesting to remark that a careful
analysis of the $\chi^2$ value for each experiment allows the
identification of potential incompatible data. This feature had
already been pointed out in the unpolarized studies by the NNPDF
Collaboration.

\FIGURE
{
  \includegraphics[scale=0.26]{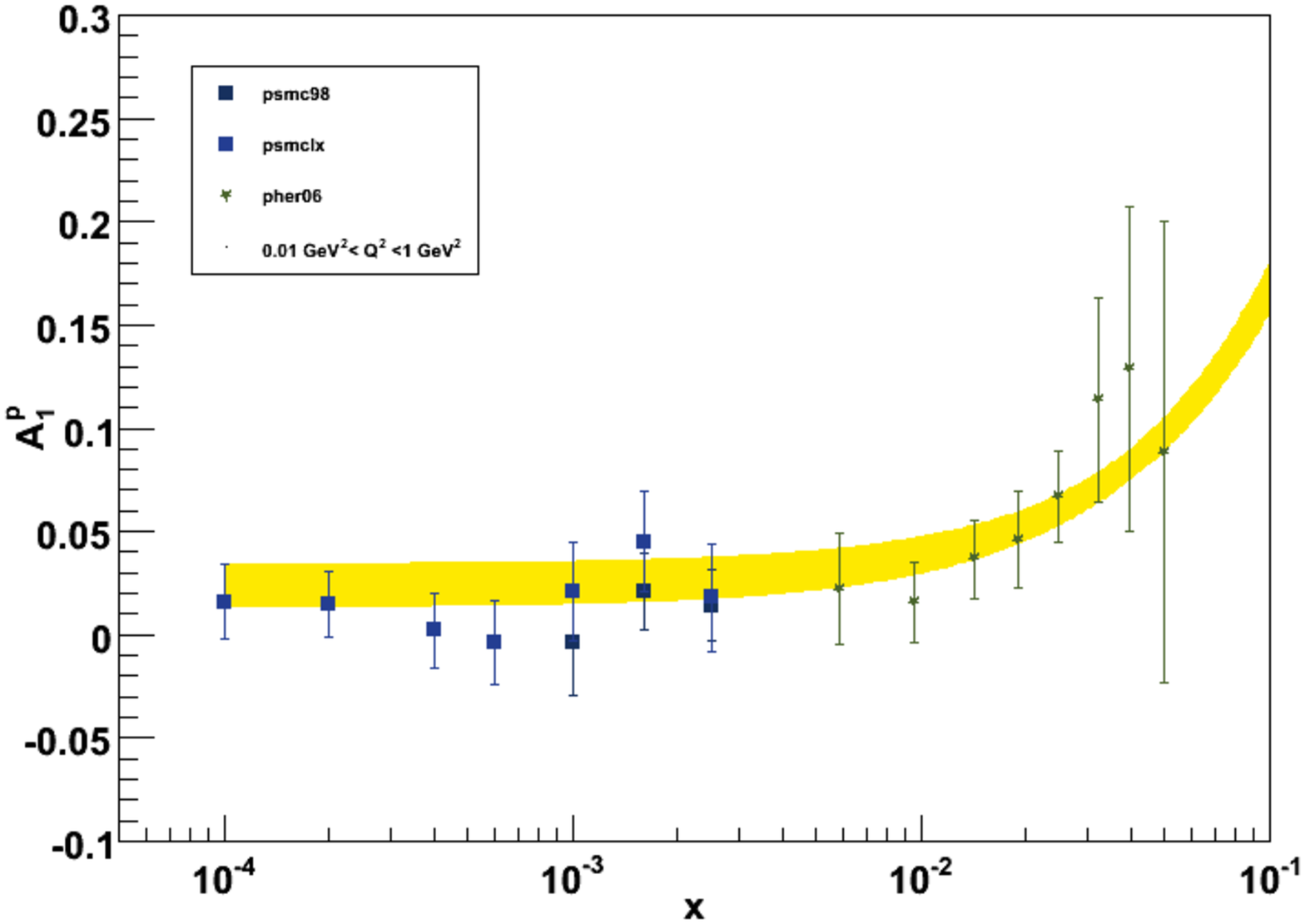}
  \includegraphics[scale=0.26]{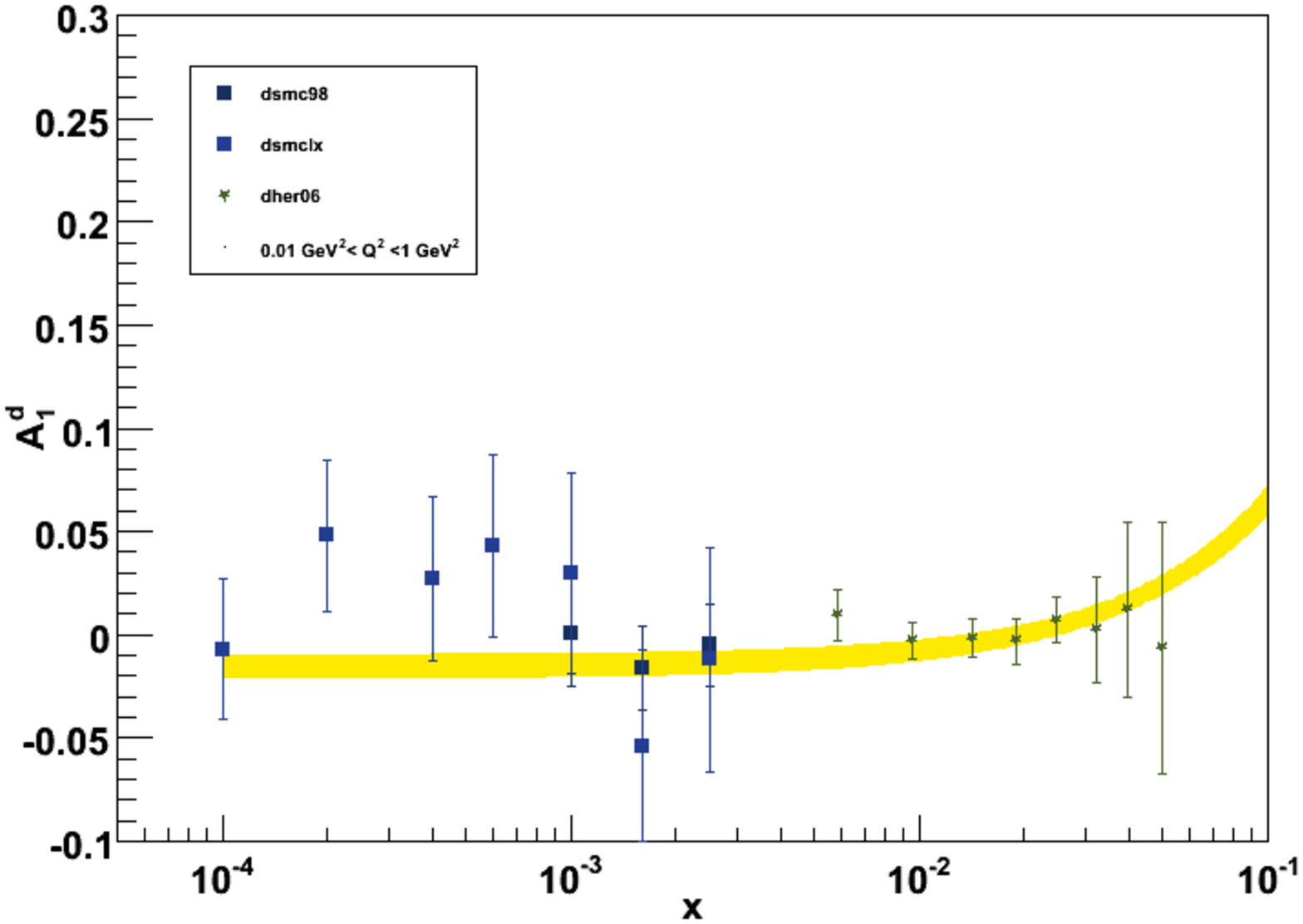}\\
  \includegraphics[scale=0.26]{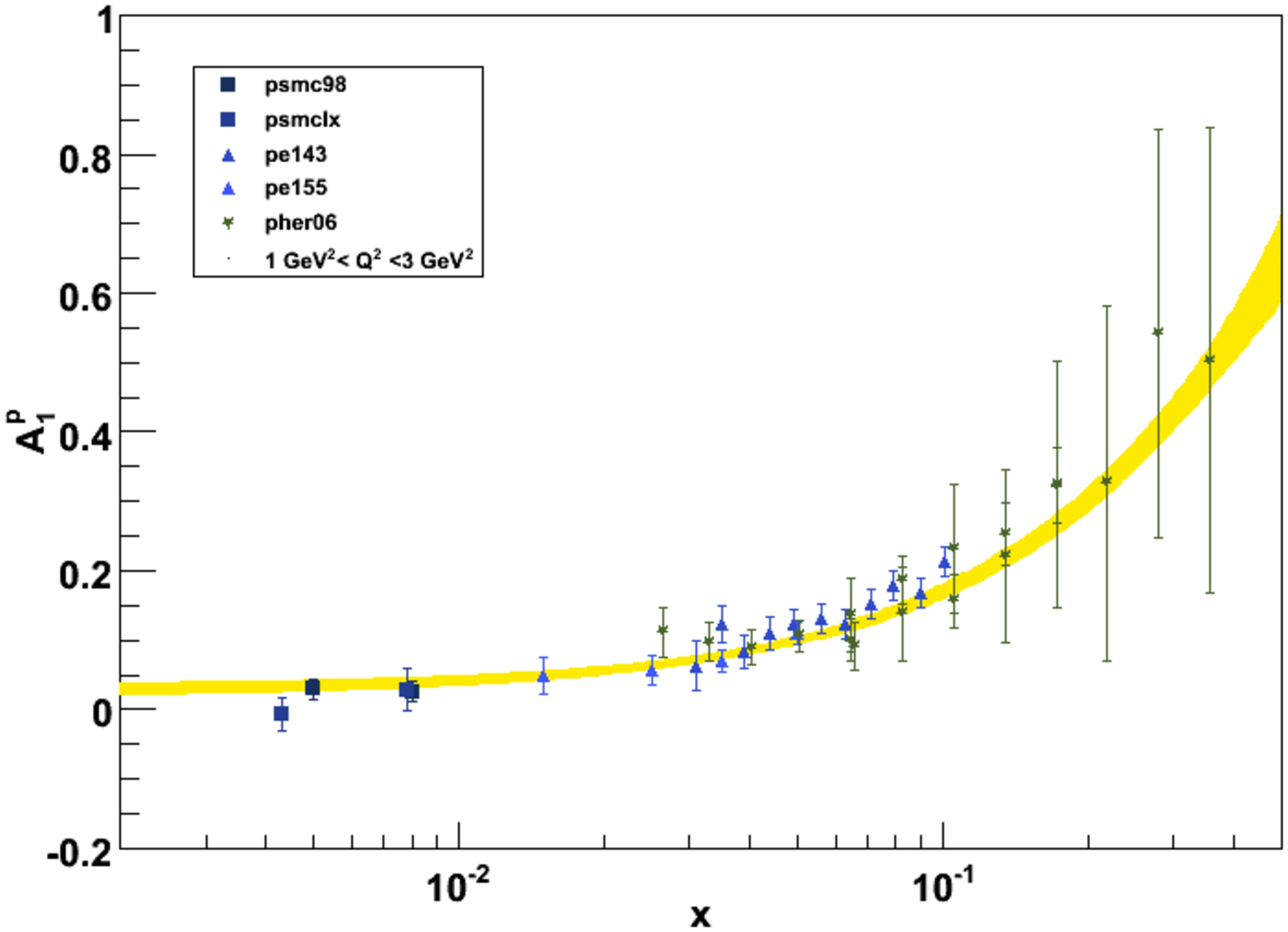}
  \includegraphics[scale=0.26]{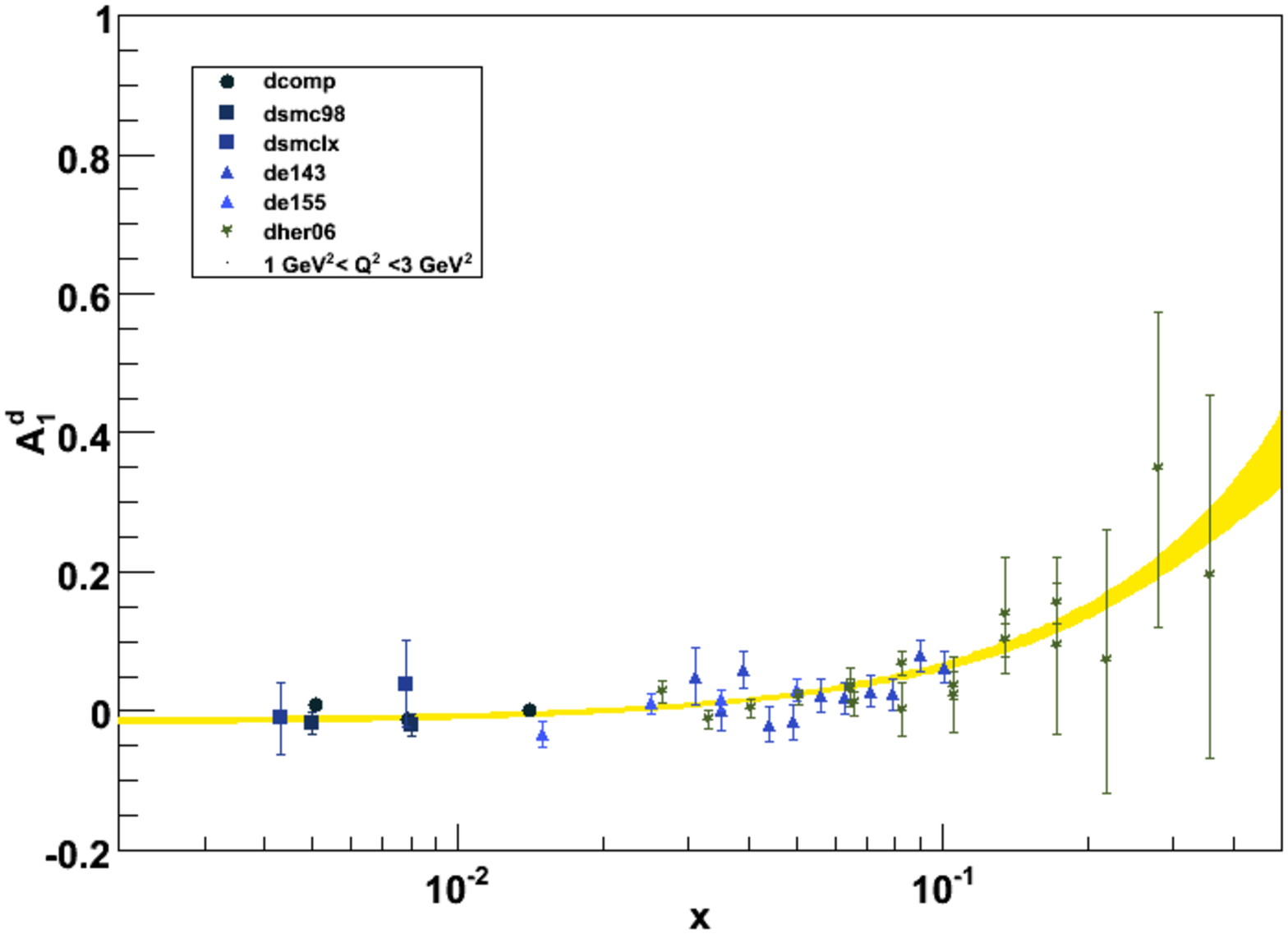}\\
  \includegraphics[scale=0.26]{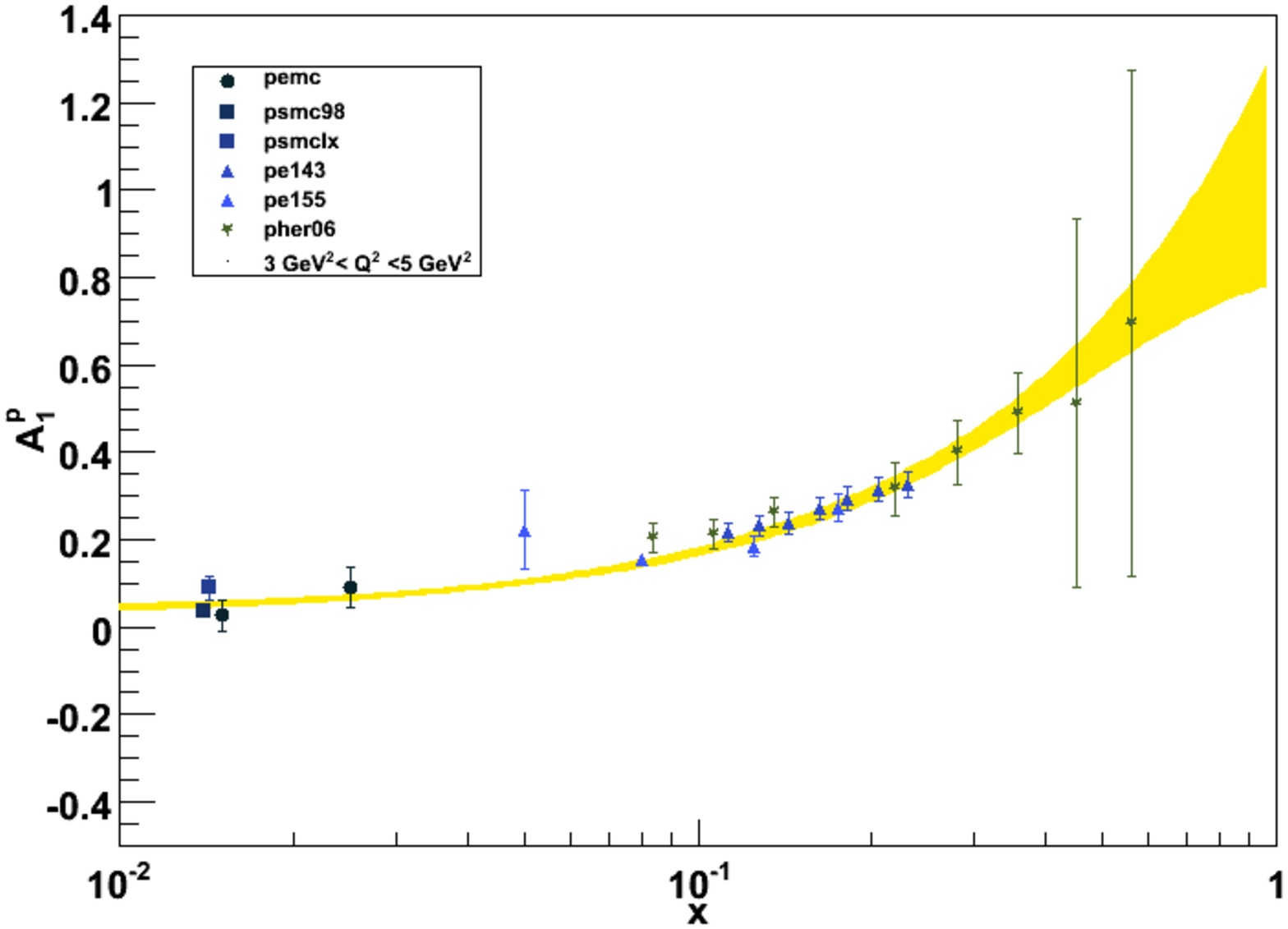}
  \includegraphics[scale=0.26]{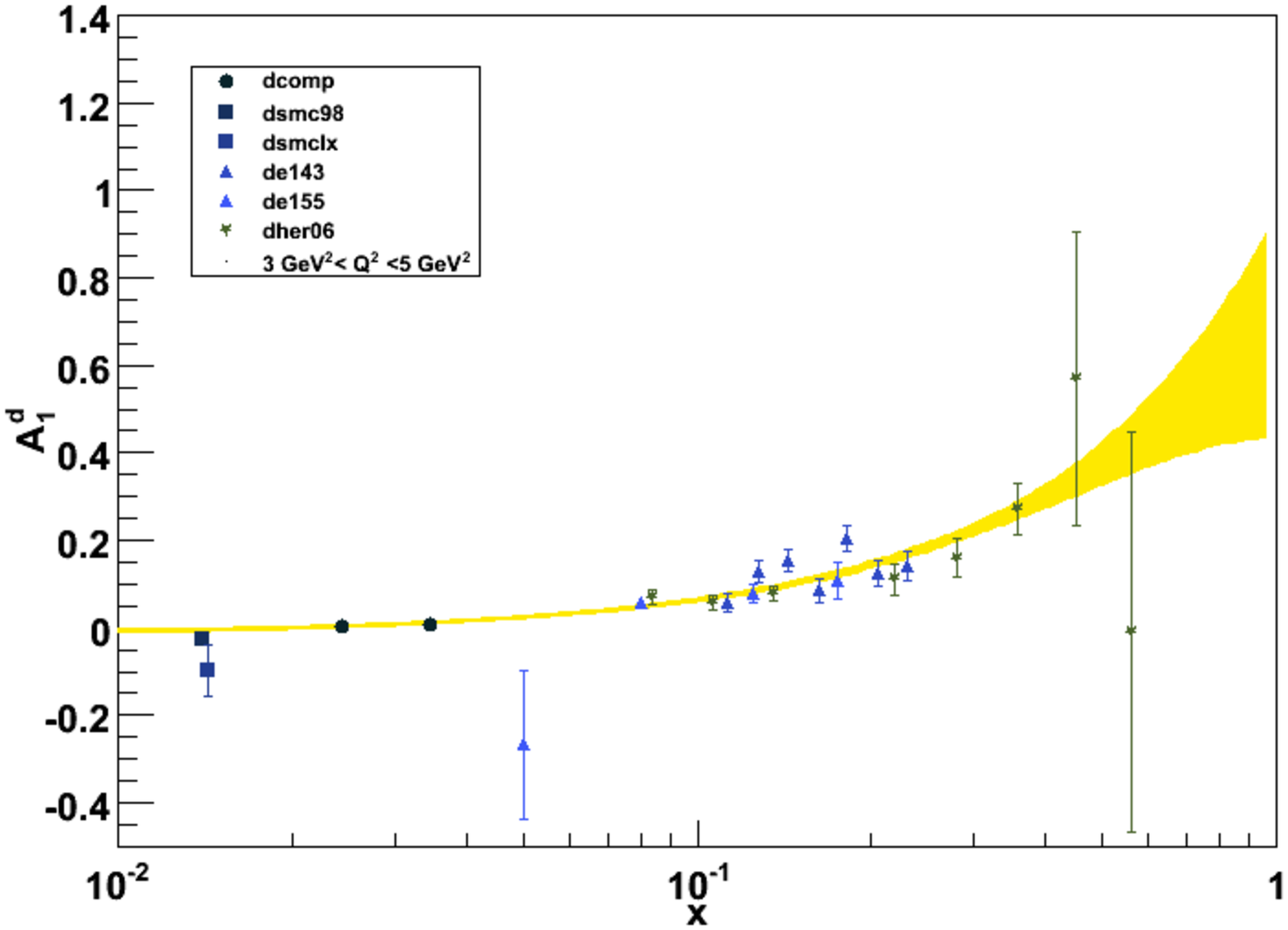}
  {\caption{The fitted asymmetries compared to proton (left) and
    deuteron (right) data for $0.01~\GeV^2< Q^2<1~\GeV^2$ (upper row),
    $1~\GeV^2< Q^2<3~\GeV^2$ (central row) and $3~\GeV^2< Q^2<5~\GeV^2$
    (lower row). In the plots $A_1$ is evaluated at the central value
    of each $Q^2$ range.}}
  \label{fig:fit_vs_data_1}
}

\FIGURE
{
  \includegraphics[scale=0.26]{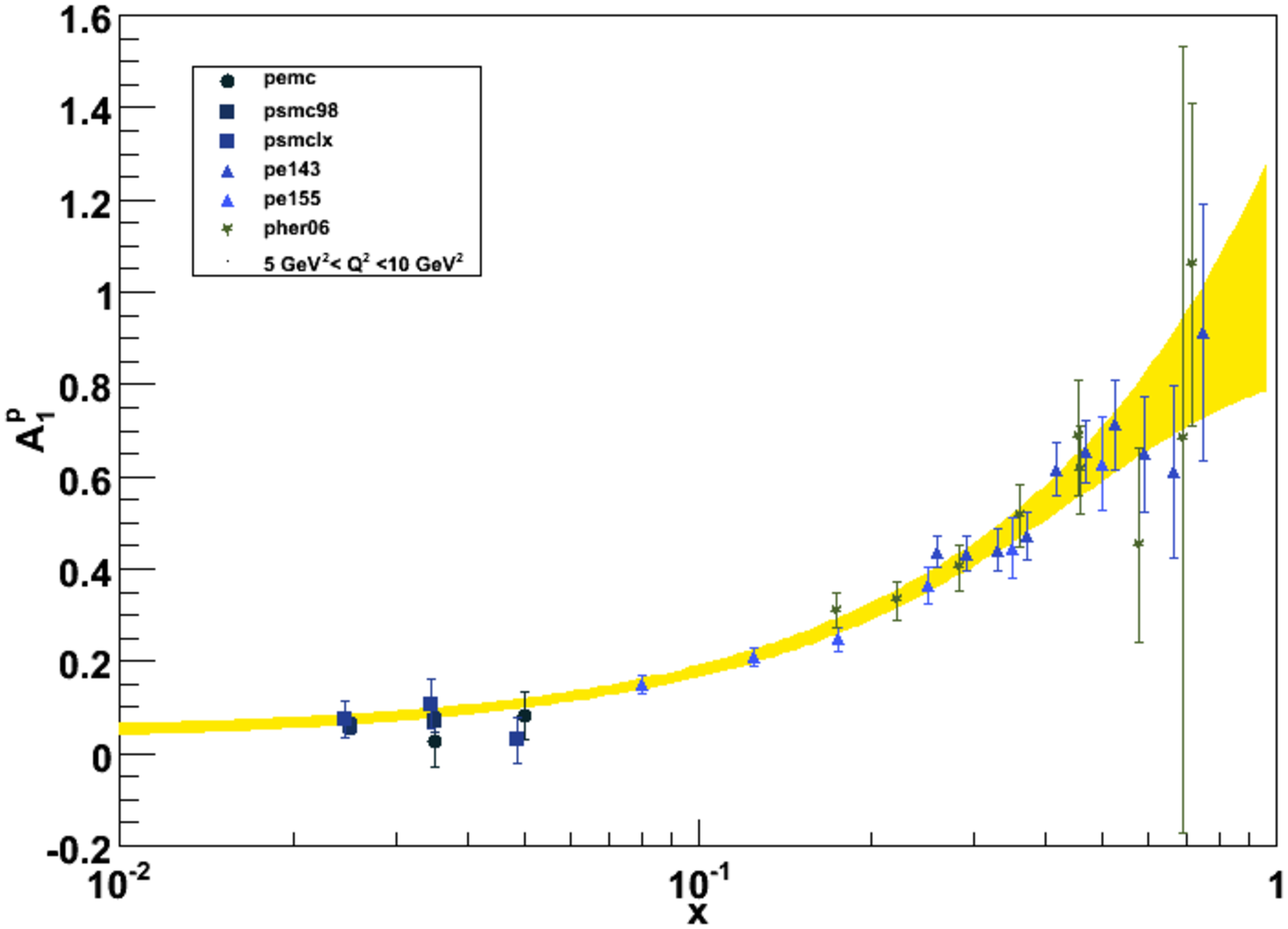}
  \includegraphics[scale=0.26]{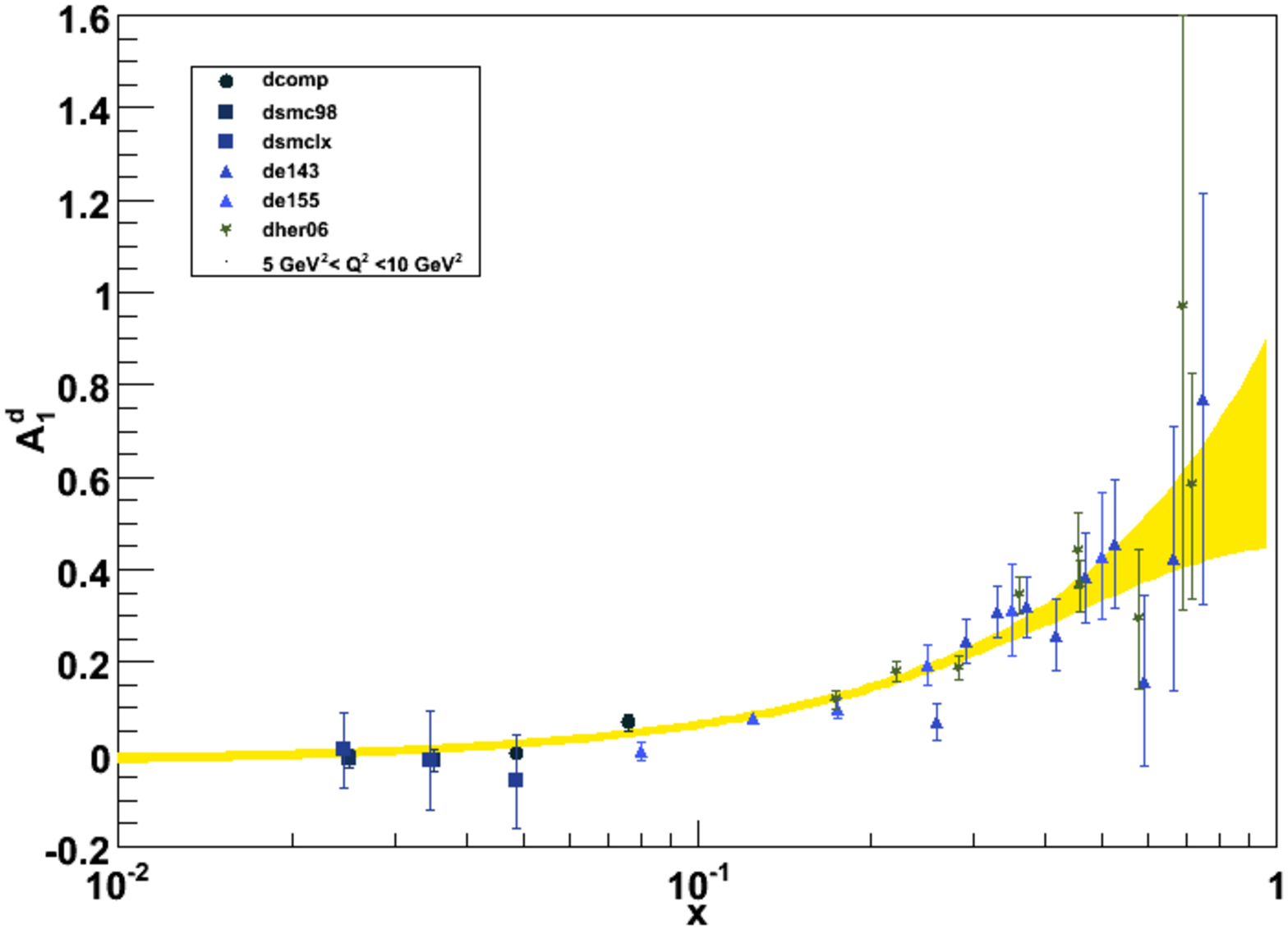}\\
  \includegraphics[scale=0.26]{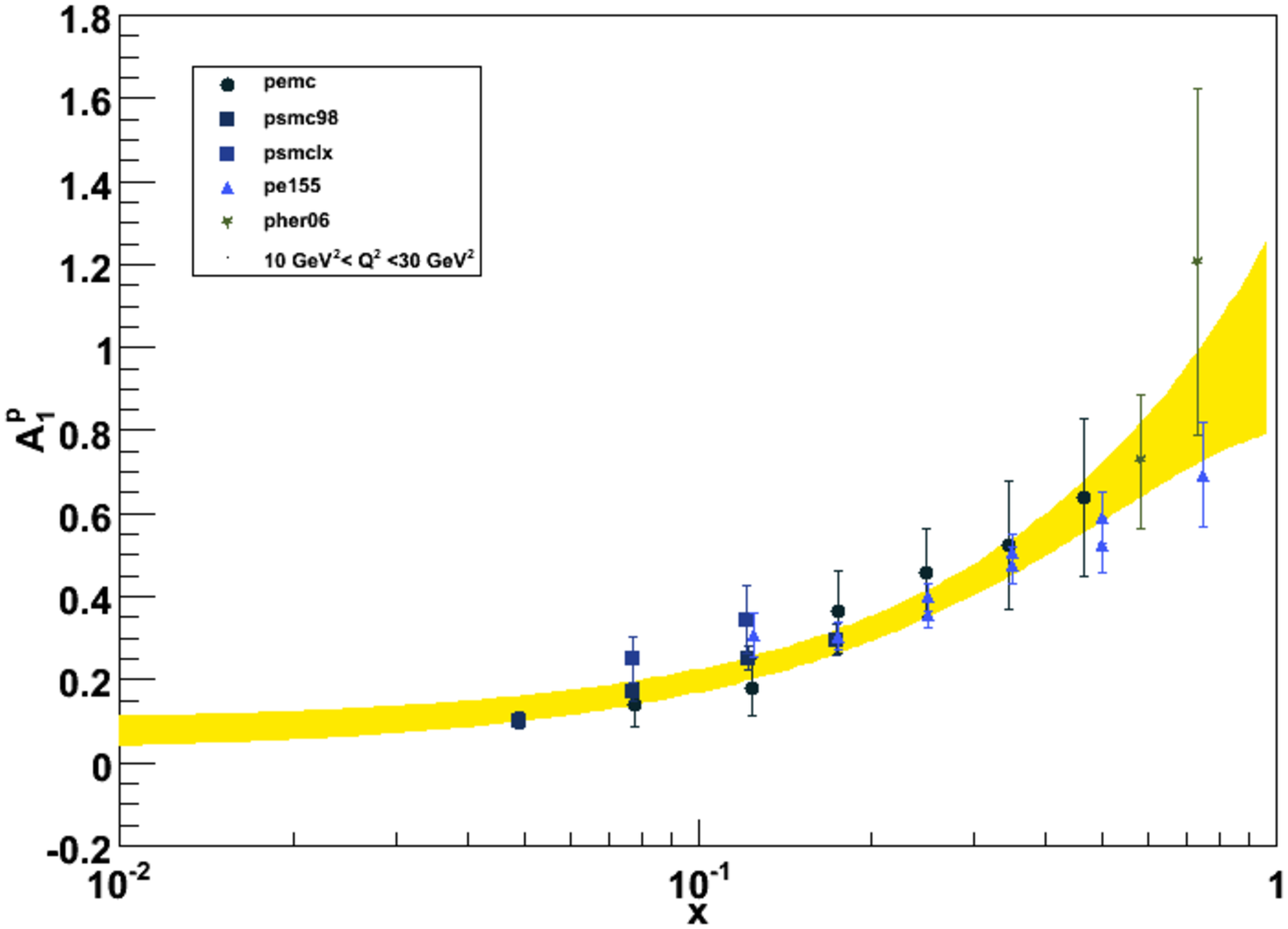}
  \includegraphics[scale=0.26]{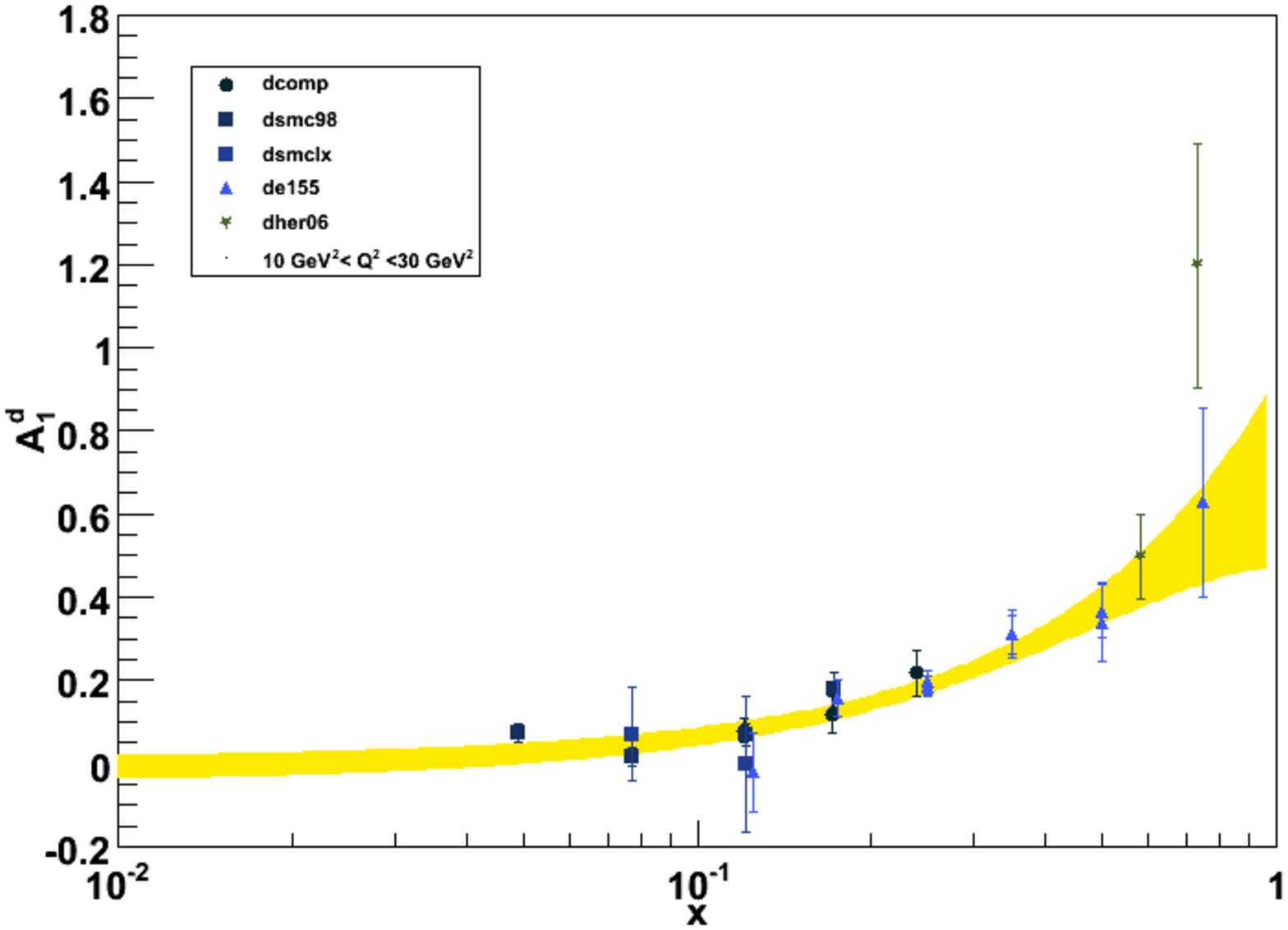}\\
  \includegraphics[scale=0.26]{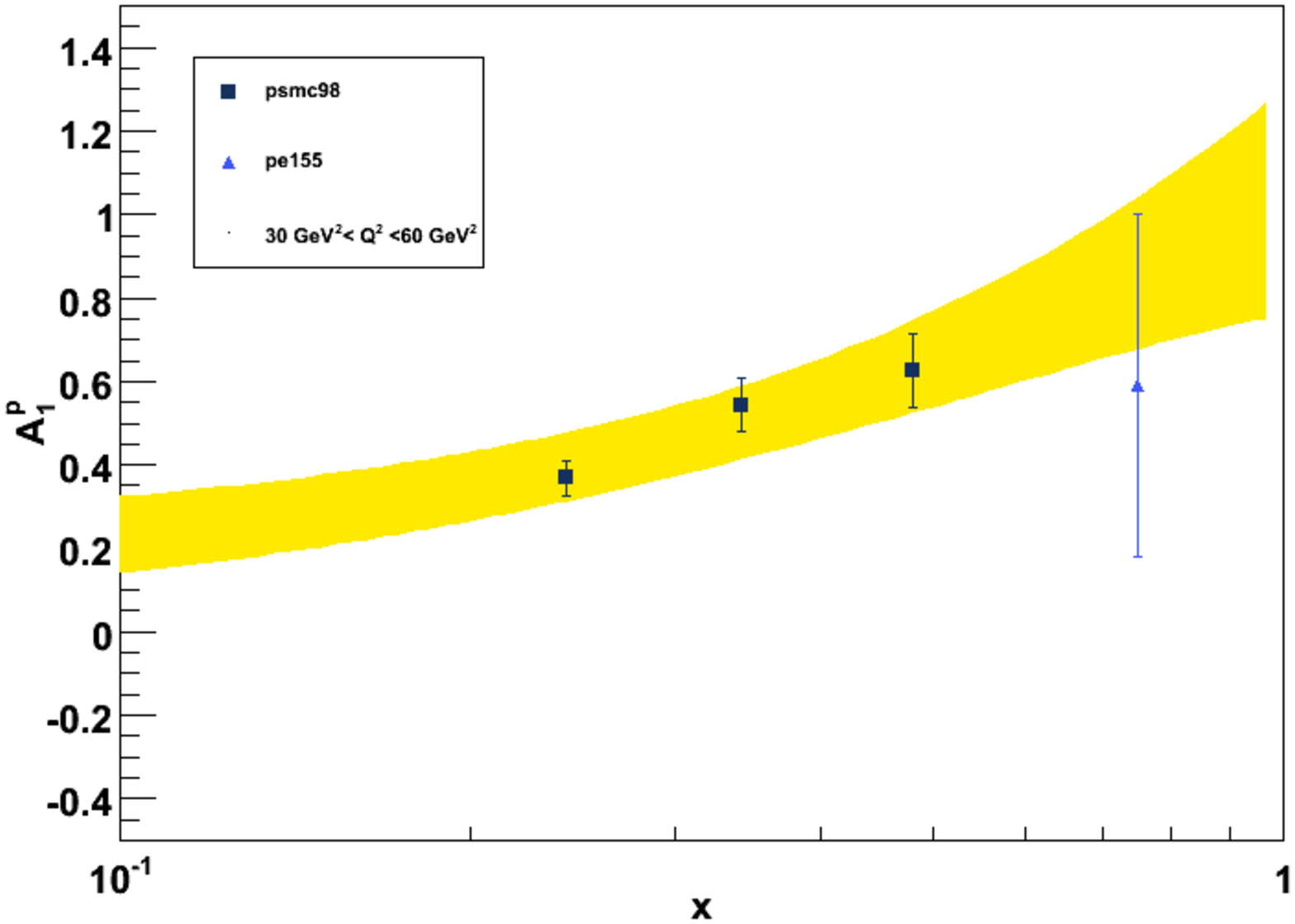}
  \includegraphics[scale=0.26]{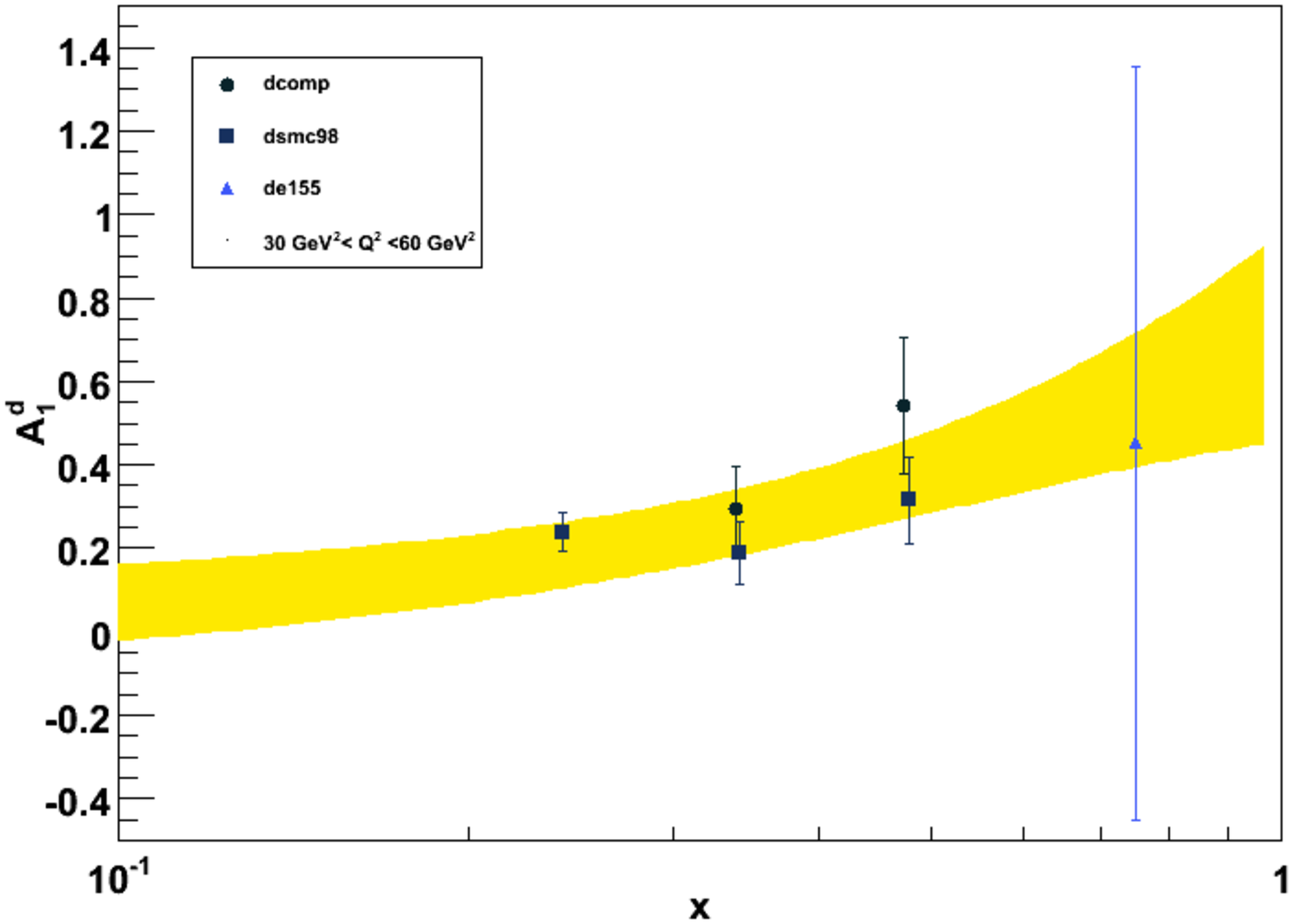}
  {\caption{The fit compared to proton (left) and deuteron (right) data
    for $5~\GeV^2< Q^2<10~\GeV^2$ (upper row), $10~\GeV^2< Q^2<30~\GeV^2$ (central row)
    and $30~\GeV^2< Q^2<60~\GeV^2$ (lower row). In the plots $A_1$ is evaluated
    at the central value of each $Q^2$ range.}}
  \label{fig:fit_vs_data_2}
}

In Tabs.~\ref{tab:self_stab}, \ref{tab:arch_stab_sml}, and
\ref{tab:arch_stab_big} we study the self--stability of the fit and the
stability against the variation of the parametrization with respect to
a smaller and a larger architecture.  To this extent we define four
different regions: one where we expect our fit to be an interpolation of 
the available data (Data region) and three where its behaviour is extrapolated
to regions of the $(x,Q^2)$-plane not covered by present data:
\begin{itemize}
\item Data: $0.01<x<0.75$ and $2~\GeV^2<Q^2<20~\GeV^2$;
\item Low-$x$: $0.0001<x<0.001$ and $2~\GeV^2<Q^2<20~\GeV^2$;
\item Low-$Q^2$: $0.2<x<0.8$ and $0.1~\GeV^2<Q^2<2~\GeV^2$;
\item High-$Q^2$: $0.2<x<0.8$ and $20~\GeV^2<Q^2<60~\GeV^2$.
\end{itemize}
We observe that all the estimators for self--stabilities are of order unity
(or smaller), meaning that different subsets within the whole ensemble of replicas 
have the same statistical features. 

When we compare our final fit to a fit performed using networks with a smaller
architecture, we notice that the the two fits are statistically equivalent. 
The same happens for the comparison with a fit done with networks with a larger architecture, 
with the only exception of the errors on the deuteron fit in the extrapolation (all distances 
are order 1.5), which show some minor instability.

\TABLE
{
  {\footnotesize
    \begin{tabular}{|l|c|c|c|c|}
      \hline
      Proton & Data  & Low-$x$ & Low-$Q^2$ & High-$Q^2$  \\
      \hline 
      $\la d[A_1]\ra$ & $ 0.963 \pm 0.011 $ & $ 0.660 \pm 0.011 $ 
      & $ 1.085 \pm 0.015 $ & $ 0.966 \pm 0.014  $ \\
      $\la d[\sigma_{A_1}]\ra$ & $ 0.840 \pm 0.006 $ & $ 0.618 \pm 0.011 $ 
      & $ 0.966 \pm 0.014 $ & $ 0.905 \pm 0.012 $ \\
      \hline
      \hline
      Deuteron & Data  & Low-$x$ & Low-$Q^2$ & High-$Q^2$  \\
      \hline 
      $\la d[A_1]\ra$ & $0.772  \pm 0.008 $ & $ 0.670 \pm 0.012 $ 
      & $ 0.804 \pm 0.012 $ & $ 0.688 \pm 0.011 $ \\
      $\la d[\sigma_{A_1}]\ra$ & $ 0.818 \pm 0.007 $ & $ 0.899 \pm 0.014 $ 
      & $ 0.730 \pm 0.011 $ & $ 0.773 \pm 0.011 $ \\
      \hline
    \end{tabular}
  }
  {\caption{Self--stability estimators evaluated with 100 replicas. The
  entries in the table show the statistical differences between
  results based on different subsets of 100 replicas randomly chosen
  in our Monte Carlo ensemble.}}
  \label{tab:self_stab}
}

\TABLE
{
  {\footnotesize
    \begin{tabular}{|l|c|c|c|c|}
      \hline
      Proton & Data  & Low-$x$ & Low-$Q^2$ & High-$Q^2$  \\
      \hline 
      $\la d[A_1]\ra$         & $ 0.952 \pm 0.010$ 
      & $ 0.792 \pm 0.014$ 
      & $ 0.859 \pm 0.012$ 
      & $ 1.295 \pm 0.016$ \\
      $\la d[\sigma_{A_1}]\ra$ & $ 1.104 \pm 0.008$ 
      & $ 1.405 \pm 0.013$ 
      & $ 0.975 \pm 0.014$ 
      & $ 1.002 \pm 0.011$ \\
      \hline
      \hline
      Deuteron & Data  & Low-$x$ & Low-$Q^2$ & High-$Q^2$  \\
      \hline 
      $\la d[A_1]\ra$         & $ 1.217 \pm 0.012 $ 
      & $ 1.302 \pm 0.012$ 
      & $ 1.324 \pm 0.017$ 
      & $ 0.703 \pm 0.010$ \\
      $\la d[\sigma_{A_1}]\ra$ & $ 0.963 \pm 0.008 $ 
      & $ 1.199 \pm 0.010$ 
      & $ 0.962 \pm 0.010 $ 
      & $ 1.689 \pm 0.019 $ \\
      \hline
    \end{tabular}
  }
  {\caption{Stability estimators for the reference fit (architecture 2-4-1)
    compared to a fit with a smaller architecture (2-3-1).}}
  \label{tab:arch_stab_sml}
}

\TABLE
{
  {\footnotesize
    \begin{tabular}{|l|c|c|c|c|}
      \hline
      Proton & Data  & Low-$x$ & Low-$Q^2$ & High-$Q^2$  \\
      \hline 
      $\la d[A_1]\ra$         
      & $ 1.076 \pm 0.012 $ 
      & $ 1.179 \pm 0.017 $ 
      & $ 0.693 \pm 0.011 $ 
      & $ 1.625 \pm 0.018$ \\
      $\la d[\sigma_{A_1}]\ra$ 
      & $ 1.258 \pm 0.010 $ 
      & $ 1.354 \pm 0.014 $ 
      & $ 0.709 \pm 0.010 $ 
      & $ 1.151 \pm 0.016 $ \\
      \hline
      \hline
      Deuteron & Data  & Low-$x$ & Low-$Q^2$ & High-$Q^2$  \\
      \hline 
      $\la d[A_1]\ra$         
      & $ 0.867 \pm 0.009 $ 
      & $ 0.794 \pm 0.012 $ 
      & $ 0.856 \pm 0.015 $ 
      & $ 0.840 \pm 0.017 $ \\
      $\la d[\sigma_{A_1}]\ra$ 
      & $ 1.460 \pm 0.009 $ 
      & $ 1.941 \pm 0.015 $ 
      & $ 1.410 \pm 0.011 $ 
      & $ 1.303 \pm 0.011 $ \\
      
      \hline
    \end{tabular}
  }
  {\caption{Stability estimators for the reference fit (architecture 2-4-1)
    compared to a fit with a larger architecture (2-5-1).}}
  \label{tab:arch_stab_big}
}

\subsection{Structure functions reconstruction}

In order to reconstruct the structure function $g_1$ from data on the
asymmetry $A_1$ as given in Eq.~(\ref{eq:A1}) some additional
assumptions are needed.  In the following we assess the impact of
our assumptions for $g_2$, $F_2$ and $R$ on the determination of first moment of
$g_1$.  These checks are done using an ensemble of 100 replicas,
which is enough to this purpose, and in a range of $x$ and
$Q^2$ which is entirely in the data region in order to avoid any
extrapolation effects. Finally, we compare our result for 1000
replicas with the sum rules obtained by experimental collaborations.

The first assumption whose impact we consider is the one on the structure 
function $g_2$, which is evaluated from the Wandzura--Wilczek relation~\cite{Wandzura:1977qf}
\bea
g_2^{WW}(x,Q^2) = - g_1 (x, Q^2) + 
\int_x^1 \frac{dy}{y} g_1(y,Q^2)\,.
\label{eq:g2WW}
\eea
Inserting this expression into Eq.~(\ref{eq:A1}) gives
\bea
g_1 (x,Q^2)= \frac{1}{1+\gamma^2}
\left(
A_1 (x, Q^2) F_1 (x, Q^2) + \gamma^2 \int_x^1 \frac{dy}{y} g_1(y,Q^2)
\right)\,,
\label{eq:g1WW}
\eea
which needs to be evaluated iteratively.
To this purpose we take the initial value $g_1 (x, Q^2)$
evaluated with $g_2 (x, Q^2)=0$, and we use
\bea
g_1^{(i_{WW})} (x,Q^2)= \frac{1}{1+\gamma^2}
\left(
A_1 (x, Q^2) F_1 (x, Q^2) + \gamma^2 \int_x^1 \frac{dy}{y} g_1^{(i_{WW}-1)}(y,Q^2)
\right)\,.
\label{eq:g1WWit}
\eea

\TABLE
{
  {\footnotesize
    \begin{tabular}{|c|r|r|r|r|}
      \hline
      Proton & $Q^2=2~\GeV^2$ & $Q^2=5~\GeV^2$ & $Q^2=10~\GeV^2$ & $Q^2=20~\GeV^2$ \\
      \hline 
      $i_{ww}=0$ & $0.1184 \pm 0.0069$ & $0.1196 \pm 0.0068$ &
      $0.1272 \pm 0.0106$ & $0.1401 \pm 0.0216$ \\
      $i_{ww}=1$ & $0.1086 \pm 0.0062$ & $0.1154 \pm 0.0066$ &
      $0.1251 \pm 0.0104$ & $0.1391 \pm 0.0215$ \\
      $i_{ww}=2$ & $0.1072 \pm 0.0061$ & $0.1152 \pm 0.0065$ &
      $0.1251 \pm 0.0104$ & $0.1391 \pm 0.0215$ \\
      \hline
      \hline
      Deuteron & $Q^2=2~\GeV^2$ & $Q^2=5~\GeV^2$ & $Q^2=10~\GeV^2$ & $Q^2=20~\GeV^2$ \\
      \hline 
      $i_{WW}=0$ & $0.0459 \pm 0.0049$ & $0.0414 \pm 0.0036$ &
      $0.0401 \pm 0.0064$ & $0.0397 \pm 0.0149$ \\
      $i_{WW}=1$ & $0.0414 \pm 0.0044$ & $0.0396 \pm 0.0035$ &
      $0.0392 \pm 0.0064$ & $0.0393 \pm 0.0149$ \\
      $i_{WW}=2$ & $0.0407 \pm 0.0042$ & $0.0395 \pm 0.0034$ &
      $0.0392 \pm 0.0064$ & $0.0393 \pm 0.0149$ \\
      \hline
    \end{tabular}
  }
  {\caption{First moment for $x$ between 0.01 and 0.75 for different number of iterations 
    of the Wandzura--Wilczek relation.}}
  \label{tab:iww}
} 
From Tab.~\ref{tab:iww} we see that for $Q^2$ values above
$2~\GeV^2$ one iteration is enough to stabilize the result for the
first moment of $g_1$ computed in the data region. For lower scales, say
$Q^2\simeq 1~\GeV^2$, at least two iterations of the Wandzura-Wilczek relation are
needed in order to obtain a stable result. In the following the index $i_{WW}$ will 
be omitted as the number of iterations used should be evident from the scale
at which the first moment of $g_1$ is evaluated.

For the unpolarized structure function $F_2$ we use the parametrization 
given in Ref.~\cite{DelDebbio:2004qj} for the proton and the one given 
in Ref.~\cite{Forte:2002fg} for the deuteron. Since these parametrizations have 
also been extracted using a Monte Carlo procedure, ensembles of replicas are available 
for $F_2$; hence the result for $g_1$ is evaluated as:
\begin{equation}
  g_1 (x,
  Q^2)=\frac{1}{N_\mathrm{rep}}\sum_{k=1}^{N_\mathrm{rep}} \left[A_1^{(k)}(x, Q^2)
    \frac{(1+\gamma^2)}{2x\left[1+R(x, Q^2)\right]} F_2^{(k)}(x,
    Q^2)+\gamma^2 g_2^{(k)}(x, Q^2)\right]\, ,
\label{eq:g1_F2ALL}
\end{equation}
which takes into account both the uncertainty on $A_1$ and the one on
$F_2$ (with $g_2^{(k)} (x, Q^2)$ we denote the expression in Eq.~(\ref{eq:g2WW}) 
evaluated for the $k$-th replica). 
Since there is no correlation between the extraction of $A_1$ and the one of
$F_2$ the replicas of $A_1$, and $F_2$ can be sampled independently.

In order to estimate the contribution of the uncertainty on $F_2$ to the uncertainty 
on $g_1$, we can recompute $g_1$ as
\begin{equation}
  g_1 (x, Q^2)=\frac{1}{N_\mathrm{rep}}\sum_{k=1}^{N_\mathrm{rep}}
  \left[A_1^{(k)}(x, Q^2) \frac{(1+\gamma^2)}{2x\left[1+R(x, Q^2)\right]}
    \langle F_2\rangle (x, Q^2)+\gamma^2 g_2^{(k)}(x, Q^2)\right]\,,
\label{eq:g1_F2AVG}
\end{equation}
where for each $k$-th replica of $A_1$ we use the averaged value
of the unpolarized structure function
\begin{equation}
\langle F_2\rangle (x, Q^2)=\frac{1}{N_\mathrm{rep}}\sum_{k=1}^{N_\mathrm{rep}}F_2^{(k)}(x, Q^2)\,.
\end{equation}
This procedure clearly freezes the fluctuations in $F_2$, which is kept fixed to its average value.
The result is given in Tab.~\ref{tab:f2}, where we see that the contribution to the uncertainty on
the first moment of $g_1$ due to $F_2$ is negligible. In the following we will always use $g_1$ as
given from Eq.~(\ref{eq:g1_F2ALL}).

\TABLE
{
  {\footnotesize
    \begin{tabular}{|l|r|r|r|}
      \hline
      Proton & $Q^2=2~\GeV^2$ & $Q^2=5~\GeV^2$ & $Q^2=20~\GeV^2$ \\
      \hline 
      Eq.~(\ref{eq:g1_F2ALL}) & $0.1086 \pm 0.0062$ & $0.1154 \pm 0.0066$ & $0.1391 \pm 0.0215$ \\
      Eq.~(\ref{eq:g1_F2AVG}) & $0.1086 \pm 0.0059$ & $0.1154 \pm 0.0064$ & $0.1391 \pm 0.0213$ \\
      \hline
      \hline
      Deuteron & $Q^2=2~\GeV^2$ & $Q^2=5~\GeV^2$ & $Q^2=20~\GeV^2$ \\
      \hline 
      Eq.~(\ref{eq:g1_F2ALL}) &  $0.0414 \pm 0.0044$ & $0.0396 \pm 0.0035$ & $0.0393 \pm 0.0149$ \\
      Eq.~(\ref{eq:g1_F2AVG}) &  $0.0414 \pm 0.0038$ & $0.0396 \pm 0.0034$ & $0.0394 \pm 0.0150$ \\
      \hline
    \end{tabular}
  }
  {\caption{First moment for $x$ between 0.01 and 0.75 with and without the error on $F_2$}}
  \label{tab:f2}
}

Finally a parametrization of $R(x, Q^2)$ is needed in order to extract
$g_1$ from $A_1$. Here we use $R_{SLAC} (x, Q^2)$ given in
Ref.~\cite{Whitlow:1990gk,Abe:1998ym}.  Such a parametrization 
provides also an error estimate, which we use to assess the impact of
$R_{SLAC}(x, Q^2)$ on the total uncertainty of the first moment of
$g_1$.  In Tab.~\ref{tab:r} we compare the sum rule evaluated with
the central value of $R_{SLAC}(x,Q^2)$ with the one obtained by taking
into account the error on $R_{SLAC}(x,Q^2)$. This is achieved by
letting $R_{SLAC}(x,Q^2)$ fluctuate within its own error in the Monte
Carlo sample; for the $k$-th replica we use:
\begin{equation}
  R_{SLAC}(x,Q^2) + r^{(k)} \Delta R_{SLAC}(x,Q^2)\,, 
\end{equation}
where $\Delta R_{SLAC}(x,Q^2)$ is the error on the parametrization,
and $r^{(k)}$ is a univariate Gaussian random number. Since
$R_{SLAC}(x,Q^2)$ is a parametrization of experimental data, we take
the error as a statistical one, with no correlation between different
replicas, and thus we use a different random number each time a value
of $R_{SLAC}(x, Q^2)$ is needed. From the results collected in Tab.~\ref{tab:r} 
we conclude that the error on $R_{SLAC}(x,Q^2)$ is also negligible.

\TABLE
{
  {\footnotesize
    \begin{tabular}{|l|r|r|r|}
      \hline
      Proton & $Q^2=2~\GeV^2$ & $Q^2=5~\GeV^2$ & $Q^2=20~\GeV^2$ \\
      \hline 
      $R_{SLAC}(x,Q^2)$ & $0.1086 \pm 0.0062$ & $0.1154 \pm 0.0066$ & $0.1391 \pm 0.0215$ \\
      $R_{SLAC}(x,Q^2)+ r^{(k)} \Delta R_{SLAC}(x,Q^2)$ 
        & $0.1086 \pm 0.0062$ & $0.1154 \pm 0.0066$ & $0.1392 \pm 0.0216$ \\
      \hline
      \hline
      Deuteron & $Q^2=2~\GeV^2$ & $Q^2=5~\GeV^2$ & $Q^2=20~\GeV^2$ \\
      \hline 
      $R_{SLAC}(x,Q^2)$ &  $0.0414 \pm 0.0044$ & $0.0396 \pm 0.0035$ & $0.0393 \pm 0.0149$ \\
      $R_{SLAC}(x,Q^2)+ r^{(k)} \Delta R_{SLAC}(x,Q^2)$ 
        &  $0.0415 \pm 0.0044$ & $0.0397 \pm 0.0035$ & $0.0395 \pm 0.0148$ \\
      \hline
    \end{tabular}
  }
  {\caption{First moment for $x$ between 0.01 and 0.75 with and without the error on $R$}}
  \label{tab:r}
}

We will now compare our results for the integral of $g_1$ at different scales and over 
different $x$ ranges obtained using our ensemble of 1000 replicas with those obtained 
by different experimental collaborations. 

\TABLE
{
  \begin{tabular}{|c|c|c|}
    \hline
    Target & SMC98 & This Analysis \\
    \hline 
    \multicolumn{3}{|c|}{$Q^2=10 \GeV^2$}\\
    \hline
    p & $0.131\pm 0.009$ & $0.139\pm 0.015$ \\
    d & $0.037\pm 0.007$ & $0.035\pm 0.011$ \\
    \hline
  \end{tabular}
  {\caption{Comparison of the proton and deuteron sum rules 
      $\left(\int_{0.003}^{0.7}dx g_1(x,Q^2)\right)$ as determined in the present analysis 
      with the results obtained by the SMC collaboration~\cite{Adeva:1998vv}.}}
  \label{tab:SMC98_SR}
}

In Tab.~\ref{tab:SMC98_SR} results for the proton and the deuteron sum rules are compared 
to the result of Ref.~\cite{Adeva:1998vv}. We observe that the results are compatible within 
errors, and that our evaluation has a larger error.

In Tab.~\ref{tab:E143_Hermes_SR} we compare our result with the ones in Refs.~\cite{Abe:1998wq},
and~\cite{Airapetian:2007mh}. First we notice that the errors are of the same size, while our
central values are systematically smaller, with a significant difference for the proton at low
$Q^2$. A substantial part of effect can be attributed to the different parametrization used for the
unpolarized structure function.  Indeed, if we evaluate the sum rule of E143 with the SMC98 $F_2^p$
parametrization \cite{Milsztajn:1990cc,Arneodo:1995cq,Adeva:1998vv}, at $Q^2=2~\GeV^2$ we obtain
$0.116\pm0.008$ which is less than one sigma away from the result in Refs.~\cite{Abe:1998wq}; the
same happens for the HERMES06 case, using the ALLM parametrization \cite{Abramowicz:1997ms} at
$Q^2=2.5~\GeV^2$ we get $0.1188\pm0.0073$.  This can be understood looking at
Fig.~\ref{fig:f2_vs_data} where we compare the different parametrizations for $F_2^p$ used in the
different analysis.

\FIGURE
{
  \includegraphics[scale=0.32]{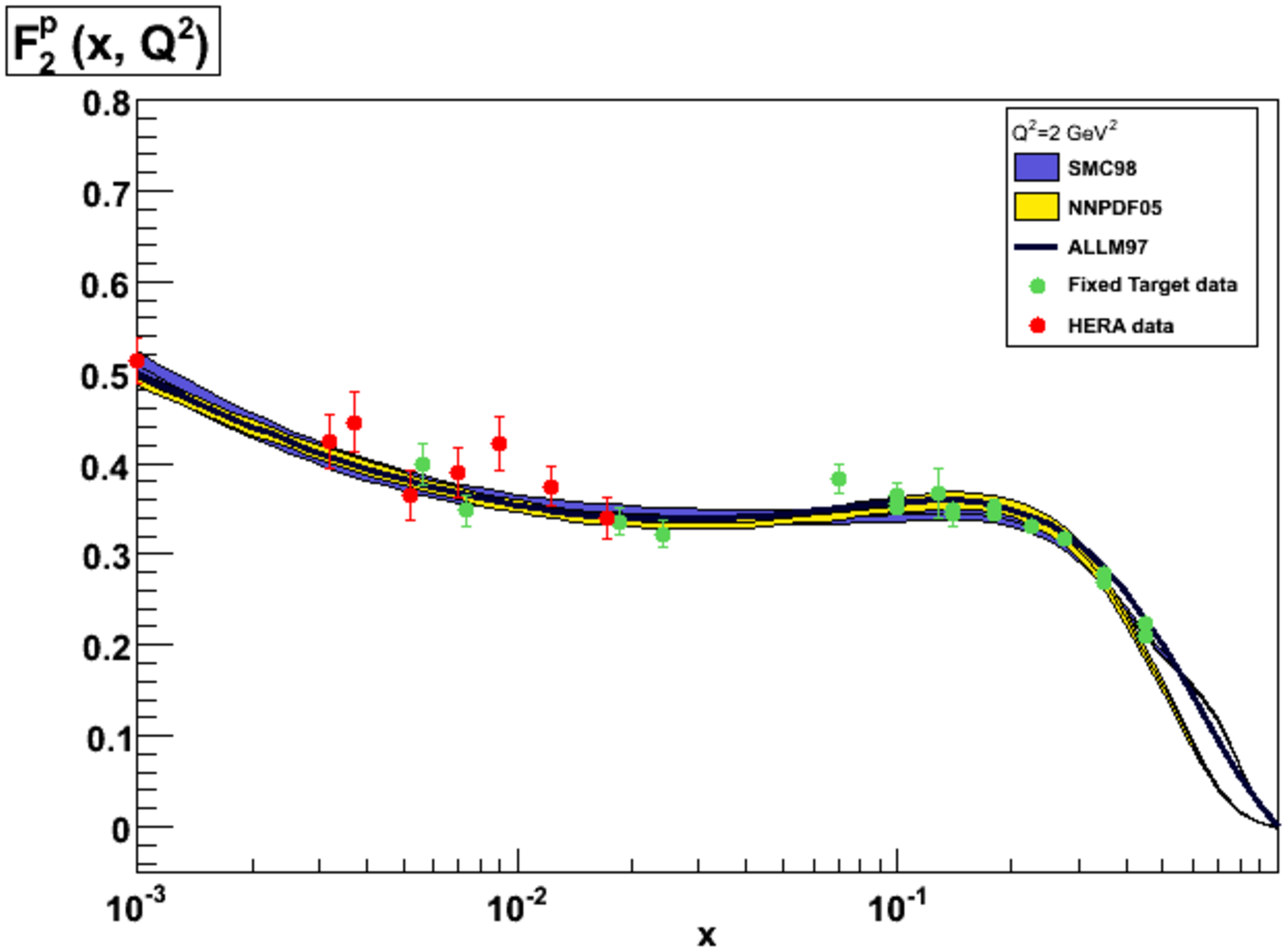}
  \includegraphics[scale=0.32]{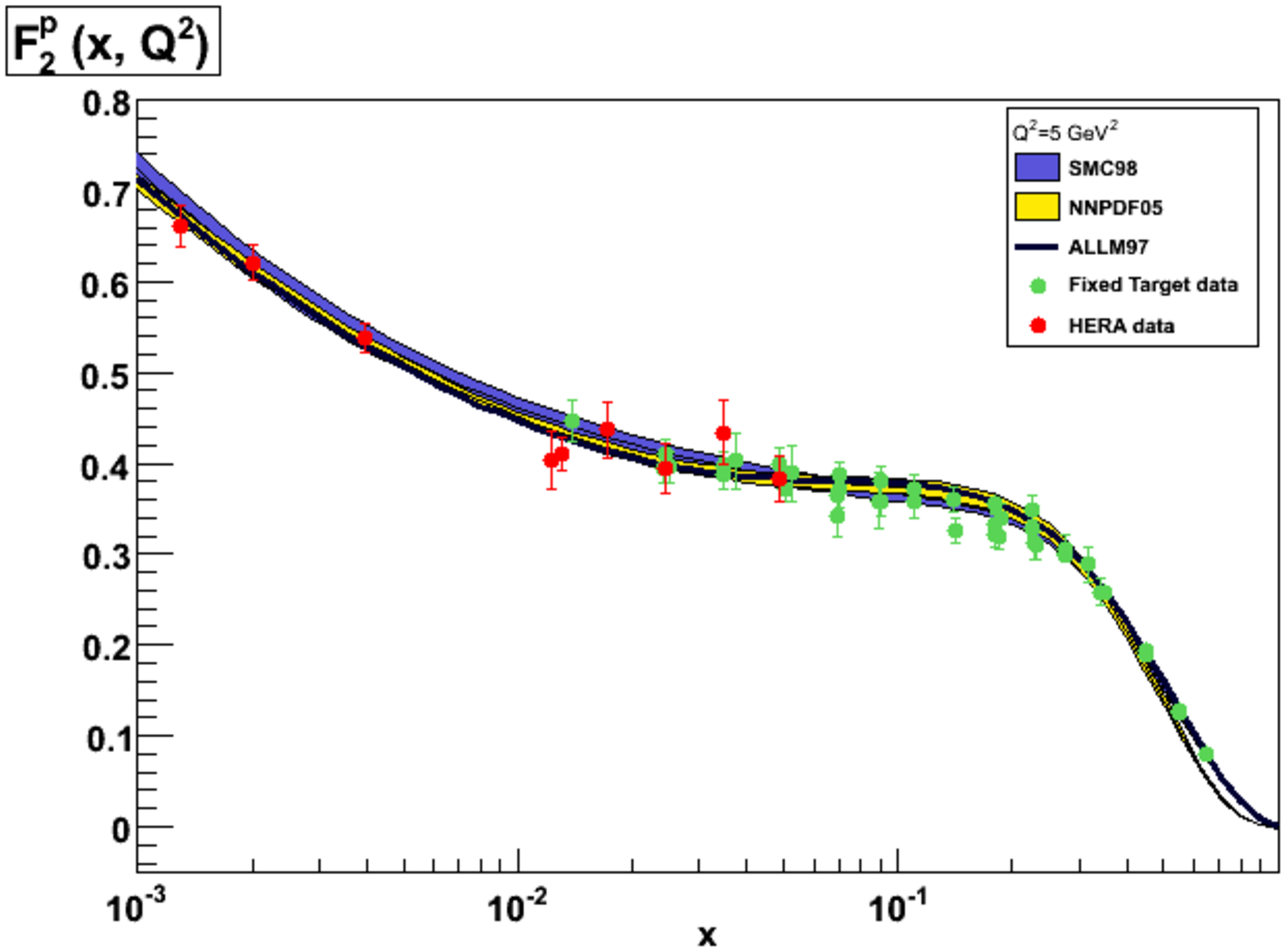}
  {\caption{Comparison of NNPDF, SMC98 and ALLM parametrizations of the unpolarized structure 
    function $F_2^p$ in the region where we evaluate the Bjorken sum rule.}}
  \label{fig:f2_vs_data}
}

It is clear that, while the different $F_2^p$ parametrizations agree in the kinematical
region covered by experimental data, they differ significantly at low-$Q^2$ in the large-$x$
region where there are no data and an extrapolation is needed.  For
the ALLM and the SMC98 parametrizations the large-$x$ behaviour is determined by 
the chosen functional form; the NNPDF parametrization
interpolates by continuity from the last experimental point to the
kinematical constrain $F_2(x=1,Q^2)=0$.  The difference among the
parametrizations is then enhanced once we multiply by the asymmetry
$A_1$ to reconstruct the polarized structure function $g_1$.

\TABLE
{\begin{footnotesize}
  \begin{tabular}{cc}
    \begin{tabular}{|c|r|r|}
      \hline
      Target & E143 & This Analysis \\
      \hline 
      \multicolumn{3}{|c|}{$Q^2=2 \GeV^2$}\\
      \hline
      p & $0.120\pm 0.007$ & $0.102\pm 0.007$ \\
      d & $0.047\pm 0.006$ & $0.042\pm 0.005$ \\
      n & $-0.022\pm 0.013$ & $-0.011\pm 0.011$ \\
      NS & $0.149\pm 0.016$ & $0.113\pm 0.016$ \\
      \hline
      \multicolumn{3}{|c|}{$Q^2=5 \GeV^2$}\\
      \hline
      p & $0.116\pm 0.007$ & $0.106\pm 0.006$ \\
      d & $0.043\pm 0.004$ & $0.040\pm 0.003$ \\
      n & $-0.025\pm 0.009$ & $-0.018\pm 0.009$ \\
      NS & $0.141\pm 0.013$ & $0.124\pm 0.014$ \\
      \hline
    \end{tabular}
    &
    \begin{tabular}{|c|r|r|}
      \hline
      Target & HERMES & This Analysis \\
      \hline 
      \multicolumn{3}{|c|}{$Q^2=2.5 \GeV^2$}\\
      \hline
      p & $0.1201\pm 0.0090$ & $0.1055\pm 0.0066$ \\
      d & $0.0428\pm 0.0035$ & $0.0416\pm 0.0043$ \\
      n & $-0.0276\pm 0.0093$ & $-0.0154\pm 0.0107$ \\
      NS & $0.1477\pm 0.0167$ & $0.1209\pm 0.0152$ \\
      \hline
      \multicolumn{3}{|c|}{$Q^2=5 \GeV^2$}\\
      \hline
      p & $0.1211\pm 0.0092$ & $0.1097\pm 0.0065$ \\
      d & $0.0436\pm 0.0035$ & $0.0407\pm 0.0033$ \\
      n & $-0.0268\pm 0.0094$ & $-0.0218\pm 0.0093$ \\
      NS & $0.1479\pm 0.0169$ & $0.1315\pm 0.0144$ \\
      \hline
    \end{tabular}
    \\
  \end{tabular}
  \end{footnotesize}
  {\caption{Comparison of the integral of $g_1$ over different $x$ ranges, at different 
      scales, as determined form the present analysis with the results obtained form by 
      E143, left pad: $\int_{0.03}^{0.8}dx g_1(x,Q^2)$  and HERMES, right pad: 
      $\int_{0.021}^{0.9}dx g_1(x,Q^2)$}}
  \label{tab:E143_Hermes_SR}
}

Since no neutron target data have been used in our fit, the neutron
structure function, $g_1^n,$ is evaluated from the proton and deuteron ones as 
\bea
g_1^n (x, Q^2)=2\frac{g_1^d (x, Q^2)}{1-1.5 \omega_D}-g_1^p (x, Q^2)\,,
\label{Dwave}
\eea 
where $\omega_D$ is the probability of the deuteron to be in the
D state; we use $\omega_D=0.05$ which covers most of the
published values \cite{omegaD}. We observe that, even if the neutron sum
rule is a pure prediction, it is compatible with other estimations.

In Fig.~\ref{fig:g1_vs_data} we show a comparison of the polarized structure function 
as extracted in this analysis to data, in the region where we evaluate the
Bjorken sum rule. The comparison shows a good agreement.

\FIGURE
{
  \includegraphics[scale=0.26]{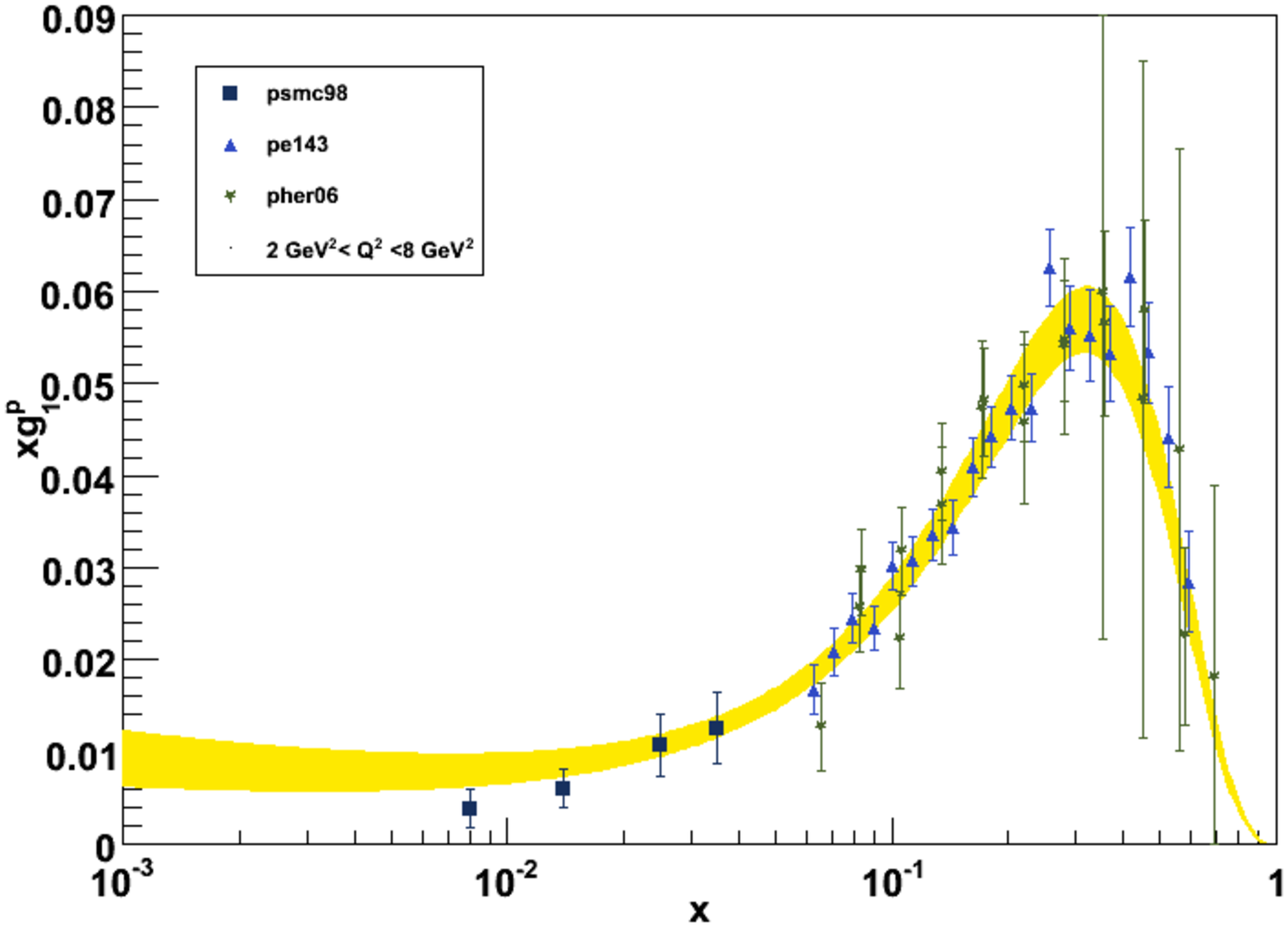}
  \includegraphics[scale=0.26]{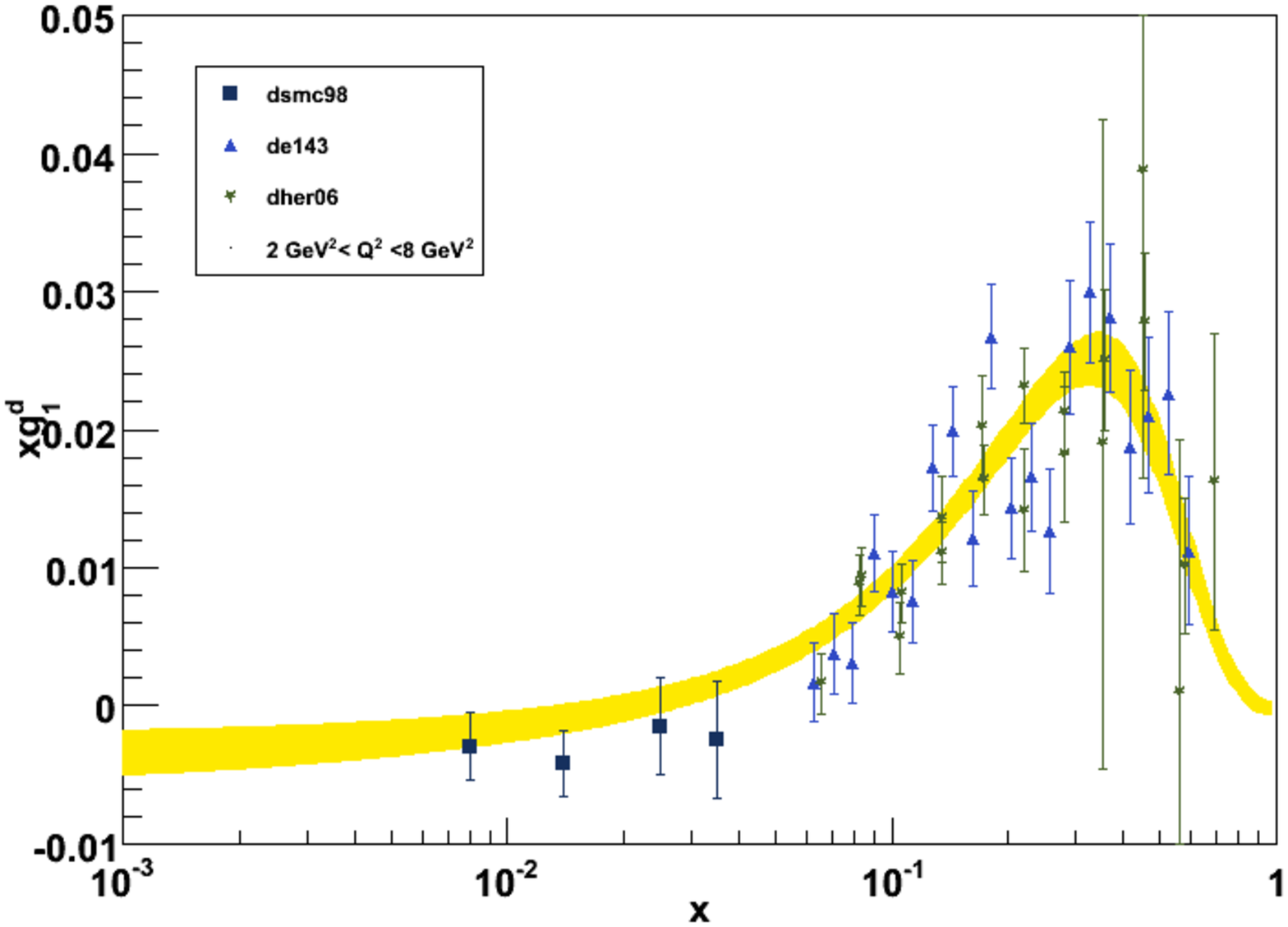}
  {\caption{Plot of the structure functions in the region where we evaluate the Bjorken sum rule.
    The fit curves are taken at $Q^2=5~\GeV^2$.}}
  \label{fig:g1_vs_data}
}

\subsection{Extraction of couplings}

In order to extract the strong coupling $\alpha_s$ and the axial coupling $g_A$ 
from the values of the Bjorken sum rule we need to extrapolate our fit in the 
Bjorken variable $x$ down to $x=0$ and up to $x=1$.

In this section we discuss the impact of these extrapolations on the extraction of the couplings,
and we assess the impact of target mass corrections. Finally we present the results we obtain for
$\alpha_s$ and $g_A$ from our fit. All checks are performed with 100 replicas, while for final
results we use the full set of 1000 replicas.

The extrapolation at large--$x$ is embedded in the parametrization of
$F_2$, as discussed in the previous section.  Therefore we do not need
any further assumption to constrain the large--$x$ behaviour. 

The low--$x$ behavior of the structure function $g_1$ is instead very weakly constrained 
by data, and the Regge behaviour is usually assumed; following Ref.~\cite{Close:1994yr} we
write:
\begin{equation}
  g_1 (x, Q^2) \simeq A\,x^b\,,
  \label{eq:g1_regge}
\end{equation}
with $0 < b <0.5$. Such an assumption requires to choose a value of $x_\mathrm{match}$ such that
for $x<x_\mathrm{match}$ the Regge behaviour is assumed to set in. The normalization factor $A$ in
Eq.~\ref{eq:g1_regge} is then determined by the matching condition
\begin{equation}
  g_1^{(\mathrm{fit})}(x_\mathrm{match}, Q^2)=A\,x_\mathrm{match}^b\,.
\end{equation}

\TABLE
{
  \begin{tabular}{|r|r|r|r|}
    \hline
    $Q^2 (\mathrm{\GeV}^2)$ & $x_\mathrm{match}$ & $\Gamma_1^{\mathrm{NS}}(Q^2)$ & Error\\
    \hline 
    1 &0.0100 &  0.12499 &  0.020989\\
    2 &0.0100 &  0.1356  &  0.018239\\
    3 &0.0100 &  0.14324 &  0.017827\\
    4 &0.0200 &  0.13847 &  0.018275\\
    5 &0.0200 &  0.14322 &  0.019021\\
    6 &0.0200 &  0.14757 &  0.020115\\
    7 &0.0200 &  0.15142 &  0.021429\\
    8 &0.0300 &  0.1458  &  0.02262\\
    9 &0.0300 &  0.14827 &  0.023859\\
    10 &0.0300 &  0.15054 &  0.025165\\
    11 &0.0300 &  0.15265 &  0.02653\\
    12 &0.0500 &  0.14111 &  0.027703\\
    13 &0.0500 &  0.14241 &  0.028749\\
    14 &0.0500 &  0.14366 &  0.029828\\
    15 &0.0500 &  0.14487 &  0.030941\\
    16 &0.0500 &  0.14604 &  0.032087\\
    17 &0.0500 &  0.14718 &  0.033267\\
    18 &0.0500 &  0.14829 &  0.03448\\
    19 &0.0800 &  0.13397 &  0.03552\\
    20 &0.0800 &  0.13461 &  0.036394\\
    \hline
  \end{tabular}
  {\caption{First moment of $g_1$ with different values of $Q^2$:
    the error on $\Gamma_1^{\mathrm{NS}}(Q^2)$ has been added a 100\% uncertainty
    on the low-$x$ extrapolation.}}
  \label{tab:xmatch} 
}

In order to choose the matching point, we fix the Regge exponent
to $0.2$ and then we proceed as in Ref.~\cite{Abbate:2005ct}: 
we evaluate the integrals at different values of $Q^2$ in the range 
$0<x<1$ for different choices of $x_\mathrm{match}$, and we look
for the minimum value of the error for each value of $Q^2$.

From the results collected in Tab.~\ref{tab:xmatch} we see that 
$x_{\mathrm{match}}$ grows as $Q^2$ gets larger. This is understood looking at 
Fig.~\ref{fig:kin_data} where we see that for larger scales the coverage of
the data moves towards higher values of $x$.

Once the matching point is been determined, in order to take into account the
uncertainty on the value of the Regge exponent and the one on the choice of the 
matching point, we randomize the Regge exponent in the range $-0.1<b<0.5$ and 
we choose the matching point to be in the range 
$x_\mathrm{match}< x < 2 x_\mathrm{match}$.

In Tab.~\ref{tab:tmc} we present the comparison for the first moment of the structure function
$g_1$ evaluated with and without the target mass correction as given in Eq.~(\ref{eq:tmc}). We
observe that the shift on the values of the moment due to the inclusion of these effects is smaller
than the experimental error even at the lowest $Q^2$.

In principle we could extract $g_A$, $\alpha_s$ and the higher-twist term by fitting
Eq.~(\ref{eq:bjsr}) evaluated from data at a given value of $Q^2$.  In practice we evaluate
$N_{Q^2}$ different moments taken at different $Q^2$ in the kinematical region where we have a good
coverage by experimental data: $2~\GeV^2 \leq Q^2 \leq 20~\GeV^2$. Indeed for $Q^2>20~\GeV^2$ the
errors on the computed moments become so large that their weighted contribution in the combination
is negligible. On the lower side of the energy range we choose to start from $Q^2=2~\GeV^2$, since
below this scale a perturbative QCD approach might not be reliable. For this reason we do not fit
the higher-twist term, but we will access its contribution by varying the lower cut in $Q^2$.

\TABLE
{
  \begin{tabular}{|c|c|c|}
    \hline
    Proton & $Q^2=1~\GeV^2$ & $Q^2=2~\GeV^2$ \\
    \hline 
    noTMC  & $0.1410 \pm 0.0109$ & $0.1264 \pm 0.0083$ \\
    wTMC   & $0.1459 \pm 0.0116$ & $0.1276 \pm 0.0086$ \\
    \hline
    Deuteron & $Q^2=1~\GeV^2$ & $Q^2=2~\GeV^2$ \\
    \hline 
    noTMC  & $0.0493 \pm 0.0141$ & $0.0350 \pm 0.0058$ \\
    wTMC   & $0.0515 \pm 0.0158$ & $0.0355 \pm 0.0059$ \\
    \hline
  \end{tabular}
  {\caption{First moment with and without TMC with $b=0.2$
    for the low-$x$ extrapolation matched at $x=0.001$.}}
  \label{tab:tmc}
}

We then proceed following the procedure described in detail in Sect. 4.3 of
Ref.~\cite{Forte:2002us}: the extraction of couplings is done by combining moments at different
values of $Q^2$ in the chosen range and fitting Eq.~(\ref{eq:bjsr}) using MINUIT
\cite{James:1975dr} where $g_A$ and $\alpha_s$ are the chosen as free parameters.  The moments at
different $Q^2$ are correlated, since they are computed using the same fitted parametrization. As
detailed in Ref.~\cite{Forte:2002us} these correlations induce numerical instabilities in the
inversion of the correlation matrix and off-diagonal instabilities due to non-diagonal elements in
the correlation matrix becoming dominant. Both these instabilities lead to unreliable results for
the extracted couplings.  
\TABLE
{
    \begin{tabular}{|c|c|c|}
      \hline
      $Q^2$ & $g_A$ & $\alpha_s (M_Z^2)$ \\
      \hline
      2+5 & 1.04 $\pm$ 0.12 & 0.126 $\pm$ 0.005 \\
      2+6 & 1.07 $\pm$ 0.16 & 0.127 $\pm$ 0.007 \\
      2+7 & 1.08 $\pm$ 0.20 & 0.128 $\pm$ 0.008 \\
      2+8 & 1.02 $\pm$ 0.21 & 0.125 $\pm$ 0.012 \\
      3+6 & 1.09 $\pm$ 0.13 & 0.131 $\pm$ 0.005 \\
      3+7 & 1.12 $\pm$ 0.17 & 0.132 $\pm$ 0.007 \\
      3+8 & 1.04 $\pm$ 0.20 & 0.127 $\pm$ 0.013 \\
      4+8 & 1.03 $\pm$ 0.17 & 0.127 $\pm$ 0.015 \\
      \hline
    \end{tabular}
  {\caption{Fits with different choices of $Q^2$.}}
  \label{tab:q2choice}
}

In order to fix the maximum value of $N_{Q^2}$ for which the extraction of the parameters 
is numerically stable and reliable, we study the error on the determination of $g_A$ and
$\alpha_s$ as we vary the number of included moments. Once we exclude moments with large 
correlations, we are left with a small number of combinations, which are showed in 
Tab.~\ref{tab:q2choice}.

The combination giving the smallest error ($Q^2=2,5~\GeV^2$), once we evaluate asymmetric errors, 
yields
\bea
  \alpha_s (M_Z^2)=0.126_{-0.009}^{+0.004}\,,
\eea
for the strong coupling, while the error on $g_A$ is found to be symmetric.

\TABLE
{
  \begin{tabular}{|c|c|c|}
    \hline
    $g_A$ & $\alpha_s (M_Z^2)$ & $k_F$ \\
    \hline
    1.02 $\pm$ 0.12 & 0.126 $\pm$ 0.006 & 0.5  \\
    1.04 $\pm$ 0.12 & 0.126 $\pm$ 0.005 & 0.75 \\
    1.01 $\pm$ 0.11 & 0.121 $\pm$ 0.006 & 1.5  \\
    1.03 $\pm$ 0.12 & 0.120 $\pm$ 0.005 & 2.0  \\
    \hline
  \end{tabular}  
  {\caption{Reference fit ($Q^2=2,5~\GeV^2$, NNLO, 1000 reps, $k_F=1$) compared with 
    variations of the factorization scale.}}
  \label{tab:factscale}
}

The only sources of theoretical uncertainty left to consider are the one due to 
the choice of factorization scale $Q=k_F m_{q}$, which we study by varying $k_F$ 
in the range $0.5<k_F<2$ and to the higher--twist contribution. The results of the variation 
of the factorization scale are shown in Tab.~\ref{tab:factscale}. To take 
into account the higher--twist contribution we take as an estimate the variation of 
the central values once the lower $Q^2$ value is moved up to $Q^2=3~GeV^2$ 
(see Tab.\ref{tab:q2choice}) and down to $Q^2=1~GeV^2$ ($g_A=0.99\pm 0.14$ and 
$\alpha_s (M_Z^2)=0.117\pm0.007$).

In conclusion, we obtain the following result for the determination of
the axial coupling $g_A$ and the strong coupling constant $\alpha_s$
\bea
g_A&=&1.04\pm 0.12 ({\rm exp.})_{-0.06}^{+0.05}({\rm theo.})=1.04\pm 0.13 ({\rm tot.}) \\ 
\alpha_s (M_Z^2)&=&0.126_{-0.009}^{+0.004}({\rm exp.})_{-0.011}^{+0.005}({\rm theo.})
                 =0.126_{-0.014}^{+0.006} ({\rm tot.})\,, \nonumber
\eea
which are compatible with previous extractions \cite{Altarelli:1996nm,Altarelli:1998nb} 
from polarized DIS and the Bjorken sum rule.

\section{Conclusions}
\label{sec:conc}

We extracted a parametrization of the spin asymmetries $A_1^{p,d}$, based on all available DIS data
using the Monte-Carlo sampling techniques and neural networks as basic interpolation tools. We
checked in the process that the statistical methods developed for the unpolarized studies by the
NNPDF Collaboration can be naturally extended to handle the new data sets considered in this
work. Our main result is an effective tool, which we used to test different assumptions needed to
reconstruct the polarized structure function $g_1$. As an example of possible applications we
compared to previous estimations of experimental sum rules, and we found that the used
parametrization for the unpolarized structure function $F_2$ can be a sizable source of error at
low values of $Q^2$.  We also performed a study of the Bjorken sum rule, and the extraction of the
axial coupling and the strong coupling, obtaining values which are compatible with previous
analysis.

It would be interesting to compare the results obtained for the Bjorken sum rule when determined
from global QCD fits to polarized DIS, SDIS and hadron-hadron collisions data, 
like the one presented in~\cite{deFlorian:2009vb}, especially once $W$ production data 
from RHIC will be included in such fits, providing an extra constraints on the light flavours 
separation.

The present study is also meant to be a first step towards the application
of the NNPDF techniques to the determination of a set of polarized PDFs with a
faithful error estimation. 

\acknowledgments

We thank S. Forte, G. Ridolfi and M. Anselmino for useful discussion and suggestions.
We are greatly indebted to R. De Vita (CLAS), L. De Nardo and A. Fantoni (HERMES), 
A. Bressan (COMPASS) and T. S. Toole (E155) for detailed information on experimental data.
The work of LDD is supported by an STFC Advanced Fellowship.
\appendix

\section{Experimental errors}
\label{sec:experr}

Experimentally, we have
\bea
A_{||}\simeq \frac{C}{f P_b P_t}\frac{N_- - N_+}{N_- + N_+}\,,
\eea
where 
\begin{itemize}
\item $C$ is a nuclear correction that depends on the material the target is made of;
\item $f$ is the dilution factor which accounts for the fact that only a fraction
of the target nucleons is polarizable;
\item $P_b$ and $P_t$ are the beam and target polarizations;
\item $N_{-(+)}$ is the number of scattered electrons/muons per incident charge
for negative (positive) beam helicity.
\end{itemize}
Thus, most of the errors quoted by experiments are normalization
errors.

\begin{itemize}
\item EMC \cite{Ashman:1989ig}: $A_2=0$ is assumed; 9.6\% overall
  normalization due to beam and target polarization; multiplicative
  errors on $R$ and $f$; additive errors on $A_2$, the false asymmetry
  $K$ and the radiative correction.
\item SMC98 \cite{Adeva:1998vv}: $A_2=0$ is assumed; multiplicative
  errors on $P_t$, $P_b$, $R$, $f$ and the polarized background
  $\Delta P_{bg}$; additive errors on $A_2$, the false asymmetry
  $\Delta A_{false}$, the radiative correction and the momentum
  resolution.
\item SMC low-$x$ \cite{Adeva:1999pa}: $A_2=0$ is assumed;
  multiplicative errors on $P_t$, $P_b$, $R$, $f$ and the polarized
  background $\Delta P_{bg}$; additive errors on $A_2$, the false
  asymmetry $\Delta A_{false}$ and the radiative correction.
\item E143 \cite{Abe:1998wq}: $g_2$ is evaluated using the
  Wandzura-Wilczeck relation 
  \bea 
  g_2^{WW}(x, Q^2)=-g_1(x, Q^2)+\int_x^1 \frac{dy}{y}g_1(x, Q^2)\,, 
  \eea 
  using and empirical fit of $g_1/F_1=ax^\alpha(1+b x+c
  x^2)(1+Cf(Q^2))$; multiplicative errors on $P_t$, $P_b$, $f$ and the
  nuclear correction $C$ which account for a total 3.7\% for the
  proton and 4.9\% for the deuteron; additive uncorrelated error on
  the radiative corrections.
\item E155 \cite{Anthony:2000fn}: $g_2$ is evaluated in the same way
of E143, but the parameters of the fitted functional form have
different values; we will add as a shift the difference between $A_1$
and $g_1/F_1$; multiplicative errors on $P_t$, $P_b$, $f$ and the
nuclear correction $C$ which account for a total 7.6\% for the proton;
additive uncorrelated error on the radiative corrections.
\item HERMES06 \cite{Airapetian:2007mh}: a parametrization for $g_2$ is fitted
  to existing data; normalizations errors of 5.2\% for the proton and
  5\% included in the systematic quoted for each data point;
  additional additive error on the parametrization used for $g_2$.
\item COMPASS \cite{Ageev:2005gh}: $A_2=0$ is assumed; multiplicative
  errors on $P_t$, $P_b$, the dilution factor $f$ and the depolarization
  factor $D$; additive errors on the false asymmetry
  and the radiative correction.
\end{itemize}


\end{document}